\documentclass{svmult}
\usepackage{makeidx}         
\usepackage{graphicx}        
\usepackage{multicol}        
\usepackage[bottom]{footmisc}
\usepackage[ansinew]{inputenc}
\usepackage{amsmath,amssymb}

\makeindex             

\begin{document}

\title*{An Agent-Based Approach to Self-Organized Production}
\author{Thomas Seidel\inst{1}, Jeanette Hartwig\inst{2}, Richard L. Sanders\inst{3}, \and Dirk Helbing\inst{4,5,6}}
\institute{Institute for Transport \& Economics\\ Dresden University of Technology, Dresden, Germany\\  \texttt{seidel@vwi.tu-dresden.de}
\and SCA Packaging Ltd., Wigan, United Kingdom\\ \texttt{jeanette.hartwig@sca.com}
\and Institute of Economic Research\\ Lund University, Lund, Sweden\\ \texttt{dick.sanders@ics.lu.se}
\and Institute for Transport \& Economics\\ Dresden University of Technology, Dresden, Germany 
\and Collegium Budapest -- Institute for Advanced Study, Budapest, Hungary
\and Department of Humanities and Social Sciences\\ ETH Zurich, Switzerland\\ \texttt{dhelbing@ethz.ch}
}

%
%
\maketitle

\begin{abstract}\mbox{ }
The chapter describes the modeling of a material handling system with the production of individual units in a scheduled order. The units represent the agents in the model and are transported in the system which is abstracted as a directed graph. Since the hindrances of units on their path to the destination can lead to inefficiencies in the production, the blockages of units are to be reduced. Therefore, the units operate in the system by means of local interactions in the conveying elements and indirect interactions based on a measure of possible hindrances. If most of the units behave cooperatively (``socially''), the blockings in the system are reduced.

A simulation based on the model shows the collective behavior of the units in the system. The transport processes in the simulation can be compared with the processes in a real plant, which gives conclusions about the consequencies for the production based on the superordinate planning.
\end{abstract}


\section{Introduction}
\label{sec:intro}

Since the world is becoming more and more complex, linear models developed in the past are increasingly failing to produce effective management tools. While the forecast of the behavior of systems such as production networks is of crucial interest for those who have to make strategic, planning, and operational decisions within a plant, current approaches are often insufficient to cope with the occuring dynamics.

\par\begin{figure}[htb]
	\centering
	\includegraphics[width=0.65\textwidth]{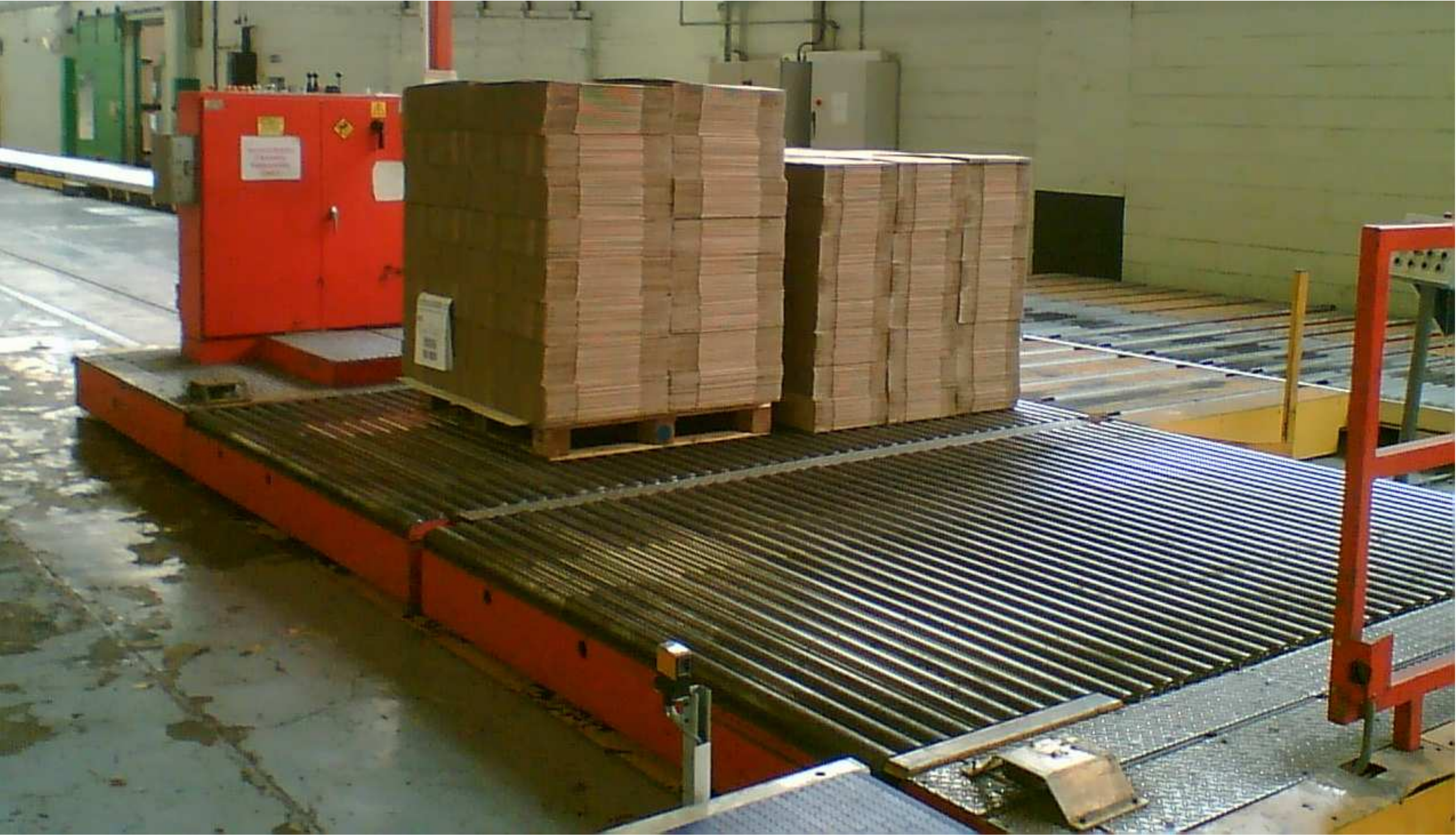}
	\hfill
	\includegraphics[width=0.32\textwidth]{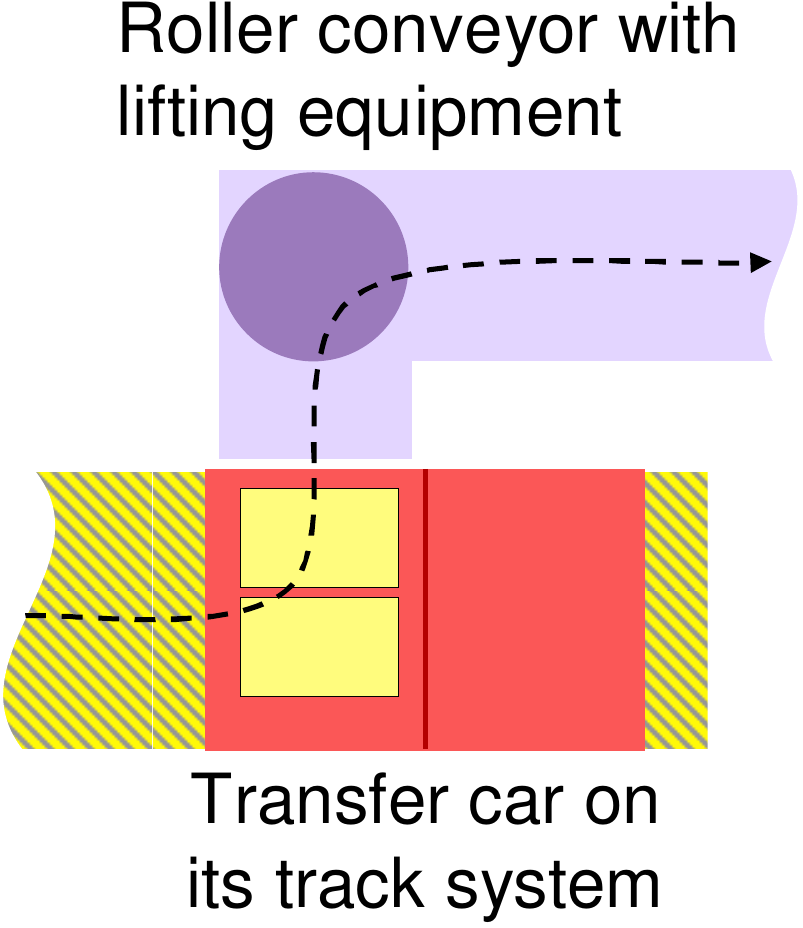}
	\caption{Left: Two stacks on a deck position of an automated guided transfer car at an intersection to a roller conveyor \cite{Hartwig2006a}. Right: Schematic representation of the element in the left figure. The change of the movement direction on the conveyor is performed by a short roller conveyor with lifting equipment (so-called chain crossover) and is marked by a circle.}
	\label{fig:tc}
\end{figure}

In the following, the modeling of complex production networks -- in particular of the packaging industry -- will be described. Figure~\ref{fig:tc} shows an element of the transport and buffer system in one of the modeled production plants and its presentation within the simulation software.

\begin{figure}[htb]
	\centering
	\includegraphics[width=0.63\textwidth]{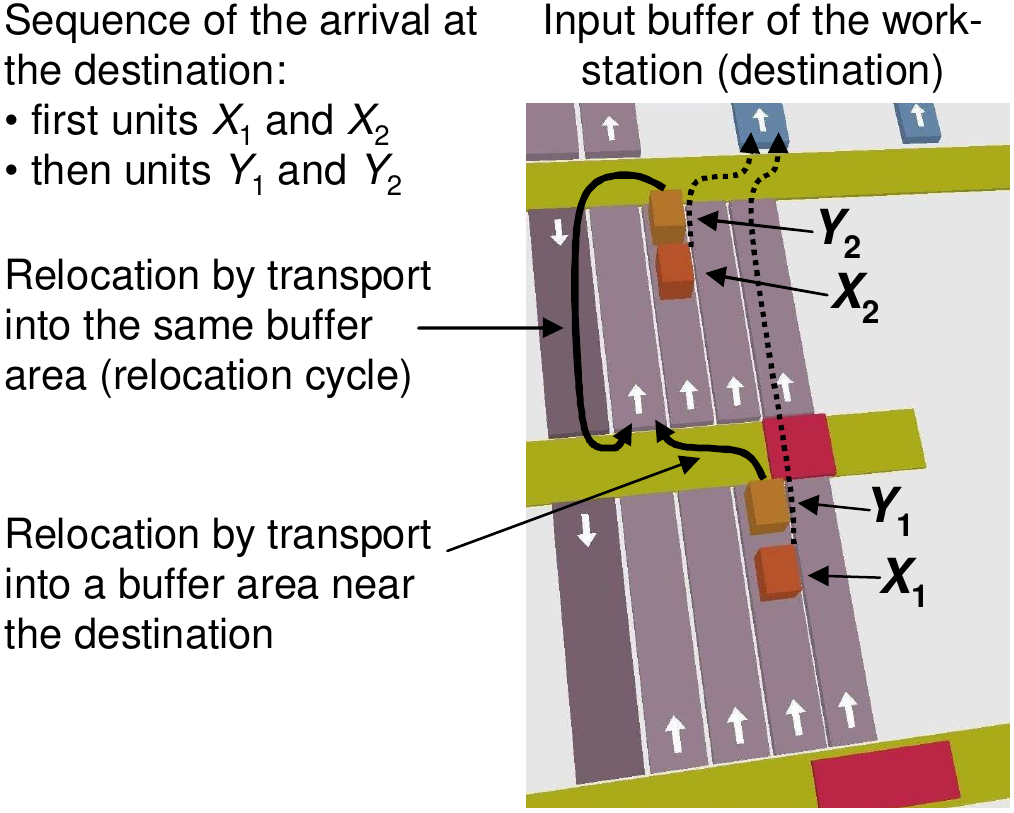}
	\caption{To meet the scheduled arrival sequence, relocations may be required to process or finish units in the right order:  In this illustration, the units $Y_1$ and $Y_2$ have to relocate, as they obstruct the units $X_1$ and $X_2$.}
	\label{fig:umlager}
\end{figure}

When modeling a general transport and buffer system within a multistage production network, one has to consider that the units (i.e. the intermediate products or work in process \cite{Hopp2000}) leave a machine in the order of their production, but are often scheduled for the next production step in a different order. Thus, a sorting of the units within the system is necessary (see Fig.~\ref{fig:umlager}). Since the planning of the production program for all machines is done centrally and in advance, the model has to take into account the given production programs. The model has to describe both, the characteristics of the transport and buffer system and the movement of the production units within the system.

The implications resulting from the required sorting of units in the system are described in Fig.~\ref{fig:umlager}. There, the workstation processes four units. The units $X_1$ and $X_2$ are authorized to enter the input buffer of the workstation first. Since $X_1$ and $X_2$ belong to the same job within the production program, i.e. they are the same type of product, and they are accepted at the workstation in any order. In our example, however, both units $X_1$ and $X_2$ are blocked by the units $Y_1$ and $Y_2$. Thus, $Y_1$ and $Y_2$ have to be removed from the lanes first. $Y_1$ will relocate into the next buffer area on its way to the destination. Unit $Y_2$ can just be transferred into the same buffer area again and, therefore, has to execute a relocation cycle.


\subsection{Problem Definition}
\label{sec:intro:problem}

The aim of this chapter is the modeling of a material handling system as an integral part of a multistage production network. Since the hindrances of units during their transport between the production stages lead to inefficiencies like shortfalls, undesired machine stops, and later completion times, blockages of units are to be reduced. Therefore, the reduction of units’ blocking is a central aspect in our model.

Due to the possibility of their mutual blocking, units have to locally \textit{interact} and \textit{react} accordingly. In order to avoid potential blockings, units also have to \textit{act in an anticipatory way} with the help of indirect interactions. This involves avoiding critical buffer areas, in which the material flow is likely to be hindered by a high buffer utilization. Thus, the aim is the design of a transport and buffer system, where local und indirect interactions help to reduce mutual hindrances of the units.

In the following, the conveying elements (e.g. roller conveyors or plastic chain conveyors) of material handling systems are called \emph{lanes}. Intersections in factories that are represented by chain transfers or turntables will generally be called \emph{turntables}. The transport and buffer system composed of lanes, turntables and transport systems (e.g. transfer cars, see Fig.~\ref{fig:tc}) will furthermore be described as mathematical graph with nodes and directed edges.

The \emph{units} to be transferred are boardstacks, production waste and auxiliary material such as ink boxes, cutting tools etc. They represent the agents in our model and operate on the nodes and the edges of the graph.

Our model is the basis for the simulation of transport processes in a plant, while a production of the units and their material handling is simulated with given production programs. The ultimate goal is to avoid hindrances of units in the real plant in order to transport them within the system to their destination on time. Simulation can help to optimize both, the transport processes given a superordinate production plan of the overall production and planning itself, if it integrates a model of the transport and buffer system.


\subsection{Organization of this Chapter}
\label{sec:intro:aufbau}

Within the first part of the chapter, the theoretical model of the transport and buffer system is introduced. The second part addresses applications of the model and examines the behavior of the units.

Section~2 provides an overview of agent-based modeling and its application to the description of transport and buffer systems. In Sect.~3, an overview of the methods used to model the production system is given, without presenting the mathematical and algorithmic details. These are provided in Ref.~\cite{Seidel2007}.

The mathematical abstraction of the functional interrelationship in a real factory is described in Sect.~4, in particular the illustration of the transport and buffer system as mathematical graph with nodes and directed edges. Afterwards, Sect.~5 summarizes the treatment of transport processes described in the previous sections.

The implementation of our model in a simulation environment is described in the second part, which begins with Sect.~6. In Sect.~7, we examine the factors contributing to a cooperative behavior of the units. Our contribution concludes with a discussion of the pros and cons of the modeling approach and evaluates the practical relevance for production systems.


\section{Relation to the Previous Literature}
\label{sec:lit}


\subsection{Material Handling Elements as Part of Production Networks}
\label{sec:lit:mat}

Production systems are generally modeled at different levels of aggregation. Frequently, the description of individual material handling elements and their interaction in larger systems is done within the framework of queuing theory \cite{Grosseschallau1984,Furmans1992,Altiok1997,Furmans2000,Arnold2005}. However, the necessary sequence of the entrance of single units into the workstation is difficult to handle. In addition, the analytical and numerical effort increases with the degree of complexity of the system and the level of detail of the elements being described.

Therefore, entire production networks or supply chains are often modeled by means of macroscopic approaches, e.g. fluid-dynamic ones \cite{Helbing2003,Helbing2004a,Armbruster2006}. These do not distinguish individuals units, which are rather modeled by event-driven simulations \cite{Ben-Naoum1995,Kouikoglou1991,DeSchutter2003}. Another modeling approach are Petri nets and the max-plus algebra \cite{Baccelli1992,Quadrat1995,Krivulin1996,Schutter2002}. One interesting feature of the latter is the possibility of analytical calculations, but a disadvantage is the effort required to adapt the description to new or modified setups.


\subsection{Agent-Based Modeling of the Transport and Buffer System}
\label{sec:lit:agent}

Agent-based modeling allows one to describe the complex interactive behavior of many individual units \cite{Darley2004,Weiss2001a,Weiss1999a,Drogoul1992}. Wooldridge \cite{Wooldridge2002} describes an ``intelligent agent'' as a computer that has the ability to perform flexible and autonomous actions in a certain environment, in order to achieve its planned goals. Agents show 

\begin{itemize}
	\item \emph{reactive} behavior in relation to the environment, 
	\item \emph{proactive} behavior (by showing initiative and acting anticipatively), and
	\item \emph{social} (e.g. cooperative) behavior.
\end{itemize}

These kinds of behavior are realized by the description of an energy and a utility function respectively, that are optimized in a distributed way. Further on, such an agent has the ability to forecast its future state or the state of the other implemented agents \cite{Darley1999}. But suitable organizational structures and communication strategies are necessary. 

A further important aspect of multi-agent systems (MAS) is the environment, in which the agents interact. The given production system consisting of the machines, connected by a transport and buffer system, constitutes the environment of the units. In accordance with the classification of Wooldridge \cite{Wooldridge2002}, the units are embedded into a dynamic and discrete production system affected by coincidences (e.g. machine breakdowns). In particular, the variability of traffic conditions on the lanes and the transfer cars is an important influence factor of the system state.

We have chosen an agent-based approach for the modeling of the units, primarily because these
agents automatically transfer their behavior to a new layout, when the factory is restructured. 
This makes an agent-based approach very flexible and easy to handle. In contrast, a classical optimization approach must be formulated for a different setup anew, which is generally quite time-consuming.

Although our modeled units show both, reactive and cooperative behavior, our agents do not act
in a fully autonomous way. They also incorporate a certain degree of central steering, which allows one to integrate our distributed control concept into a hierarchical optimization and production planning. Note that it would, in principle, also be possible to implement a centrally controlled buffer operating strategy steering the units \cite[p.~494]{Gudehus2000}, based on hierarchical optimization. For this, one would usually start with the central determination of the optimal arrival sequence at the workstations, which requires to solve a scheduling problem \cite{Kumar1995,Lu1991,Kumar1990,Perkins1989}. Next, one could describe the movement of the units to a workstation as a vehicle routing problem (VRP) \cite{Domschke1997a}, which is another central optimization approach. As a result, fixed routes and time windows would be assigned to each unit (see the overview in Refs. \cite{Braeysy2005} and \cite{Moehring2005,Lau2003,Desrosiers1984}). Intersecting flows (for example at turntables and chain crossovers) could be steered in the same way as in the control of traffic lights \cite[pp.~128]{Mehlhorn1995}. However, at least for online control, a purely central description of the entire system as VRP would be unsuited due to the enormous complexity of the solution space, the numerically demanding search algorithm, and the considerable variability of real production processes.

In contrast to the classical optimization approach sketched above, we will in the following propose a distributed control of material flows \cite{Laemmer2006,Helbing2005c}. This decentralized approach fits perfectly to an agent-based approach and has a greater flexibility, robustness, and performance under largely variable conditions such as the ones observed in many production systems with unexpected machine breakdowns, last minute orders, and other surprises.

Note that, besides the units, the material handling elements of the transport and buffer systems can also be treated as agents. Although the lanes do not perform independent actions, they are involved in the interaction processes with the units. The transfer cars are service agents, which react to requests from the units and relocate them. The transport systems, to which the transfer cars are assigned, can again be understood as VRP. An optimal driving strategy for the cars can be found for a given time window by solving this VRP \cite{Psaraftis1980,Jaw1986,Gendreau1998,Ascheuer2000,Nikolakopoulou2004}.


\subsection{Interactions as Basis for ``Social'' Behavior}
\label{sec:lit:inter}

In MAS, ``social'' behavior of the agents is of substantial importance. In the last years, promising metaheuristics have been developed for the description of interactions leading to cooperative behavior \cite{Braeysy2005,Blum2003,Talbi2002}: Ant Colony Optimization (ACO) is an interesting multi-agent approach to the modeling of transportation problems \cite{Dorigo2004,Bonabeau2002,Bonabeau2000,Kube2000,Corne1999,Dorigo1992}. ACO is motivated by social, self-organizing insects \cite{Camazine2001,Theraulaz1999,Bonabeau1998,Theraulaz1995a,Theraulaz1995}. In ACO, the agents show the behavior of ants and move along the edges of a mathematical graph. The goal is the creation of efficient routes between the nodes with the help of distributed optimization \cite{Bertelle2002,Stilwell1993}. Indirect communication between the social insects is facilitated by \textit{pheromones}, which are deposited along the edges \cite{Dorigo2004,Dorigo2000}. The feedback via the variable pheromone concentration can trigger an emergent collective behavior of the insects \cite{Peters2005,Dussutour2004}.

Another approach is inspired by investigations of interactive pedestrian behavior. The basis of these considerations are models of self-driven many-particle systems \cite{Helbing2001b}. A pedestrian regulates his or her speed and moves purposefully. In addition, all individuals react to other participants according to attractive or repulsive interaction effects (``social forces''), changing their actual speed and direction of motion. Investigations have shown that, for medium pedestrian densities, lanes consisting of pedestrians with the same desired walking direction are formed \cite{Helbing2001c}. In ``panic situations'', however, the increased excitement of the pedestrians may generate intermittent mutual obstructions: Large noise amplitudes lead to a ``freezing by heating effect'' \cite{Helbing2005a, Helbing2000b, Helbing2002, Helbing2000a}.


\subsection{Ingredients and Properties of the Modeled Transport and Buffer System}
\label{sec:lit:ing}

As indicated before, our model of the transport and buffer system is based on an agent-based concept. The agents represent units (or material handling elements), who can interact in direct and indirect ways with each other. The direct interaction takes place locally between the units in a lane and is described by the ``interaction component'' (see Sect.~\ref{sec:over:ini}).

For the indirect interactions, a hindrance coefficient is formulated, which has a similar function as pheromones for social insects. It describes the effect of possible hindrances in a lane (see Sect.~\ref{sec:over:hind}). For the path finding, we use a network algorithm that was developed to solve shortest path problems \cite{Ahuja1993}. Our algorithm considers the hindrance coefficients of the lanes and the temporal restriction given by the scheduled arrival at the workstation (see Sect.~\ref{sec:over:search}). The resulting indirect interactions between the units via the hindrance coefficients (stigmergy) support a decentralized optimization, steering the flow of units between the workstations similarly to self-organized traffic light control \cite{Laemmer2006}. In connection with the indirect interaction principle, this path finding induces a movement based on the current situation in the plant. This procedure leads to proactive behavior of the units. 

The problems of sorting and obstruction avoidance (see Sect.~\ref{sec:res:blockade}) are tackled by the combination of a distributed and a centralized approach (which may, however, be reformulated in terms of a decentralized approach as well): On the central level, all units heading for a single workstation are brought into the correct order (according to the production schedule) by a classical sorting procedure.  The units receive their time of scheduled arrival from the assigned destination (e.g. workstation). On the local level, the sorting takes place via reactive local interactions (see Sect.~\ref{sec:over:select}). If a unit is blocked by another unit in the lane due to a wrong order, the blocking unit is informed and will often decide to leave the lane, thereby clearing the congestion (see ``interaction component'' in Sect.~\ref{sec:over:ini}).

Thus, the model contains decentralized interactions that enable flexible adjustments to the current situation in the plant. By local interactions, hindrances can be successfully resolved or even avoided. However, the scheduled arrival at the workstation is centrally determined by the production program, which tries to reach a high throughput at low costs (i.e. little waste and few machine setups).


\section{Overview of Model Ingredients}
\label{sec:over}

In the following subsections, we will describe our model, which can delineate arbitrary networks of material handling elements. The following questions must be answered:
\begin{itemize}
	\item How do units find their paths in \textit{arbitrary} networks, so that they arrive at the destination at the right time? 
	\item How do the units \textit{interact} with each other, so that they obstruct each other as little as possible and arrive in the \textit{correct order} at the destination? 
	\item How is the future action of the units determined by the goal of avoiding mutual hindrances?
\end{itemize}

\begin{figure}[htb]
	\centering
		\includegraphics[width=0.7\textwidth]{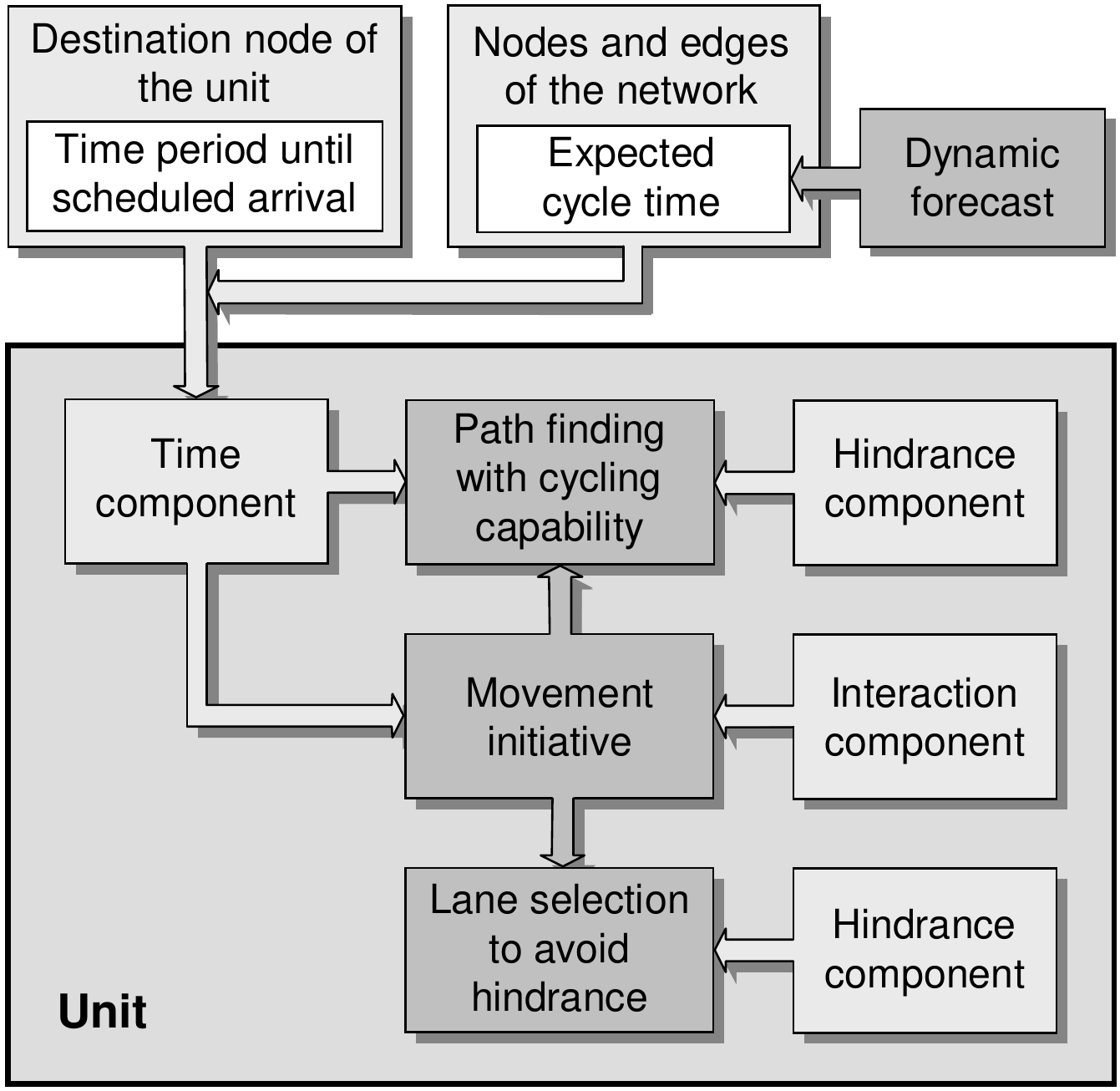}
	\caption{Overview of modeling methods and their interdependencies.}
	\label{fig:grob}
\end{figure}

\noindent
Within our model, the plant layout is represented by a mathematical graph with nodes and directed edges, which is described in Sect.~\ref{sec:steps} in more detail.

In general, a transport and buffer system consists of lanes (e.g. conveying elements like roller conveyors) that can be loaded or unloaded by transport systems like automated guided vehicles or transfer cars on tracks. The lanes are usually equipped with engines and are automatically steered by photo sensors. Most of the lanes carry the material in only one direction (``first in, first out''). Therefore, the material flow of the lanes is assumed to be directed in our model.

As in real plants, several lanes that are connected to the same transport systems and follow the same direction of the material flow are combined to a buffer area. The buffer areas represent the nodes of the graph and are linked by the transfer systems. An edge of the graph corresponds to a transport connection of two buffer areas. Due to the directedness of the material flow within the lanes, the edges are directed as well.

The model of the transport and buffer system consists of the following procedures, whose interdependencies are illustrated in Fig.~\ref{fig:grob}:

\begin{enumerate}
	\item \textbf{Dynamic Forecast of the Expected Cycle Time and Estimation of the Possible Hindrances in Lanes}: After determining a fast and hindrance-minimal route in the system by path finding, both the expected cycle times and the possible hindrances of each unit in the lanes are estimated.
	\item \textbf{Path Finding with Cycling Capability and Automatic Determination of the Hindrance-Minimal Buffer}: A deviation from the fastest path is permitted by an informed search strategy \cite{Russell2002}, in order to allow moving to a hindrance-minimal buffer area. The path finding routine has the particular capability to generate cycles in the route.
	\item \textbf{Movement Initiative}: The unit basically decides about its transport and buffering in the lane according to its own priority, but  it considers requests of other units in the same lane to move away.
	\item \textbf{Selection of the Next Lane to Avoid Hindrances}: During the relocation into the next buffer area, the following lane is selected, taking into account obstructions which, at a later time, may result from entering that lane.
\end{enumerate}

\noindent
Essential aspects and the results of the procedures are described in the following sections 
(for details see Appendix C in Ref.~\cite{Seidel2007}).


\subsection{Dynamic Forecast of the Expected Cycle Time of a Lane}
\label{sec:over:ct}

In this section, we describe the operation of the lanes and the basic elements of the transport and buffer system. The lanes are supposed to transport the units and, at the same time, provide a buffering possibility. Therefore, it is \textit{practically} impossible to distinguish between transport and buffering in the system. However, when \textit{modeling} the entire cycle time, a distinction between transport and buffering time is necessary. Whether a unit in a lane is buffered or transported is decided on the basis of the superordinate buffer operating strategy\footnote{Buffer operating strategies can be divided in allocation strategies and movement strategies \cite{Gudehus2000}. The strategy of uniform distribution is a typical allocation strategy and distributes the units uniformly over all buffer areas. The strategy ``first in, first out'' is a movement strategy and performs the entrance and exit of units in the same order.} or the temporal urgency of the unit to arrive at its destination on time (see Sect.~\ref{sec:over:ini}).

\par\begin{figure}[htb]
	\centering
		\includegraphics[width=1.0\textwidth]{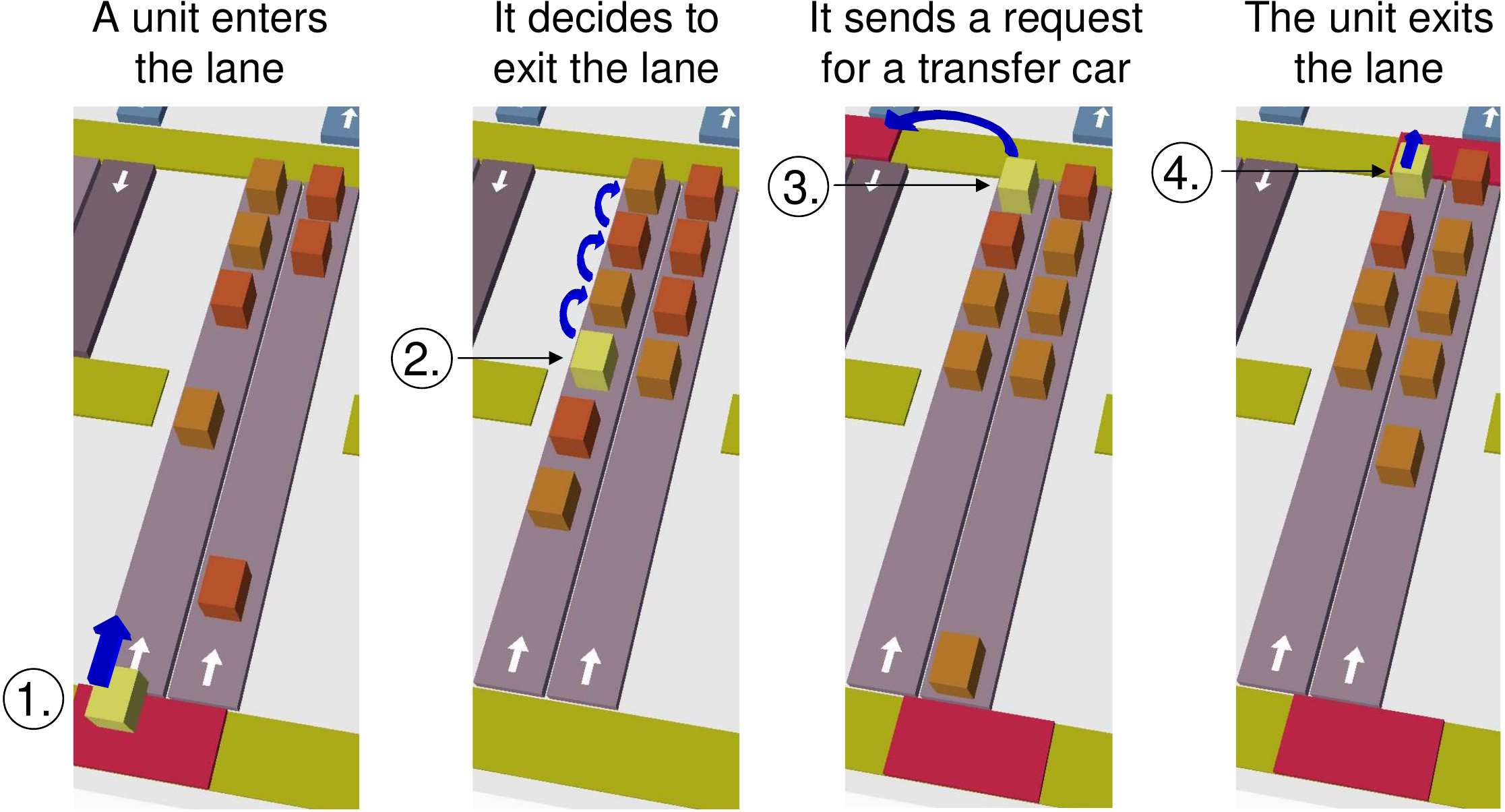}
	\caption{Subsequent actions of a unit from the arrival in a lane to its removal. The information flow is symbolized by arrows.}
	\label{fig:lager}
\end{figure}

Transport within a buffer lane takes place in four steps (see Fig.~\ref{fig:lager}): 
\begin{enumerate}
	\item The transfer car selects a lane for the unit based on the current system conditions, the expected cycle time and the possible hindrances in the alternative lanes. The unit then \textit{enters} that lane and is carried to its last free position to wait there. 
	\item At a certain time, the unit triggers the transport to the next buffer area by the ``movement initiative''. If the unit is blocked, it reports its priority to the blocking unit, this one to the next etc., until the priority message reaches the first unblocked unit. Then, all units will try to free up the lane, given their priorities are lower.
	\item If the unit with the highest priority is no longer blocked after the exiting of the hindering units, it releases a \textit{request} for the transfer car. 
	\item As soon as the transfer car is available, it loads the unit and transfers it to the next lane (\textit{exit of the unit}).
\end{enumerate}

In principle, the time period of a unit in a lane is determined by different influence factors, which include 
\begin{itemize}
	\item actual transport,
	\item buffering, 
	\item the blockage by other units in the lane, and
	\item the time until the requested transport to the next lane takes place. 
\end{itemize}
The forecast of the cycle time starts by assuming transport with no waiting (buffering). The actual duration of buffering as well as the best suitable buffering location are determined by a path finding algorithm with cycling capability.

The transport time $t(l)$ on a lane $l$ depends on its level of occupancy. The expected cycle time of a unit entering the lane later on is estimated via the cycle times of the units that entered and exited previously. The estimation of the cycle time takes place by a forecast 
\begin{itemize}
	\item of the period during which the unit is blocked by other units in the lane (the duration from the second to the third step) and 
	\item of the period beginning with the request for the transfer car (duration from the third step to the fourth step).
\end{itemize}
We have performed this forecast with an adjusted method of double exponential smoothing (ES2), which is an extension of the classical single exponential smoothing (ES1) \cite[p.~60]{Huettner1986}. The ES2 can make a trend prediction (instead of a simple smoothing performed by the ES1). 

In the exponential smoothing algorithms, the observed values $t_n$ with $n\in\mathbb{N}$ and the smoothing factor $\alpha\in(0,1)$ are given. The ES1 $\tilde{t}_n^{(1)}$ calculates 
\begin{equation} \label{eq:es1}
	\tilde{t}_n^{(1)} = \tilde{t}_{n-1}^{(1)} + \alpha\cdot \bigl(t_n-\tilde{t}_{n-1}^{(1)}\bigr)
\end{equation}
with the forecast value $\tilde{t}_{n+1}=\tilde{t}_n^{(1)}$. By means of formula (\ref{eq:es1}) the ES2 $\tilde{t}_n^{(2)}$ is calculated by
\begin{equation} \label{eq:es2}
	\tilde{t}_n^{(2)} = \tilde{t}_{n-1}^{(2)} + 
		\alpha\cdot \bigl(\tilde{t}_n^{(1)}-\tilde{t}_{n-1}^{(2)}\bigr).
\end{equation}
The forecast is then
\[
	\tilde{t}_{n+1} = \tilde{t}_n^{(1)} +
	  \frac{1}{1-\alpha}\bigl(\tilde{t}_n^{(1)}-\tilde{t}_n^{(2)}\bigr).
\]
Since ES2 is a trend function, it can predict unrealistically small or even negative values when the observed values decrease. We have, therefore, adjusted the procedure in a way that takes into account the minimum possible value $t_{\min}$ by
\[
	\tilde{t}_{n+1} \ge t_{\min}.
\]
Therefore the adjustment of the ES2 has to fulfill the condition
\[
	 \tilde{t}_n^{(1)} + \frac{1}{1-\alpha}\bigl(\tilde{t}_n^{(1)}-\tilde{t}_n^\text{korr}\bigr) \ge t_{\min}.
\]
Considering formula (\ref{eq:es2}), this leads to
\[
	\tilde{t}_n^\text{korr} = \min\Bigl\{ \tilde{t}_{n-1}^\text{korr} + 
		\alpha\cdot \bigl(\tilde{t}_n^{(1)}-\tilde{t}_{n-1}^\text{korr}\bigr),\;
		\tilde{t}_n^{(1)} + (1-\alpha)\cdot\bigl(\tilde{t}_n^{(1)}-t_{\min}\bigr)	\Bigr\}.
\]
Finally, the forecast value is determined by
\begin{equation} \label{eq:es2_pred}
	\tilde{t}_{n+1} = \tilde{t}_n^{(1)} +
	  \frac{1}{1-\alpha}\bigl(\tilde{t}_n^{(1)}-\tilde{t}_n^\text{korr}\bigr).
\end{equation}
The prediction (\ref{eq:es2_pred}) is performed every given time step or upon entering or exiting the lane.


\subsection{Dynamic Forecast of Possible Hindrances in a Lane}
\label{sec:over:hind}

A unit has to consider also the possible hindrances in the lanes, while searching for a suitable path through the system. On the one hand, hindrances can be considered as blockage, if units that are already buffered in the lane block the exit of another unit. On the other hand, the units in the lane can be removed by force, in order to allow for an unhindered transport of a newly entering unit. 
This will be often connected with relocation cycles. 

\begin{figure}[htb]
	\centering
	\includegraphics[width=1.0\textwidth]{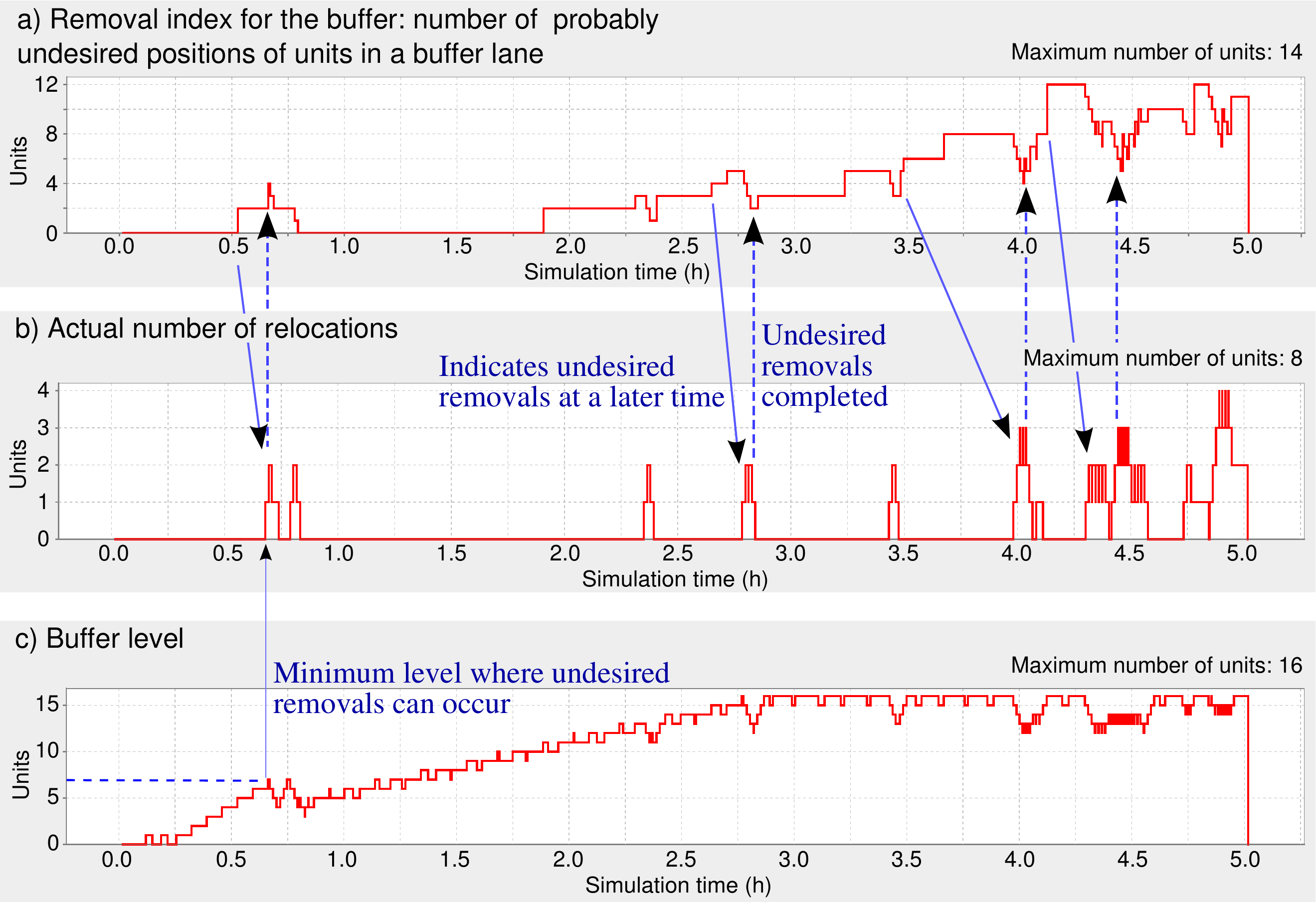}
	\caption{Comparison of the relocation cycles with the removal index and number of units for a buffer area with 2 lanes and 8 positions each. a) Removal index. b) Amount of transfer lane necessary for the relocation cycles. The causal influences are represented by arrows. c) Buffer level.}
	\label{fig:abh1}
\end{figure}

For the estimation of possible hindrances from the point of view of a new unit entering a lane, the sequence of all units in the lane is compared with the scheduled order of exits. The so-called \textit{removal index $R(l)$} counts the number of undesired positions of units in a lane $l$, resulting from a comparison of both sequences (see Fig.~\ref{fig:abh1}). 

Let us assume $n$ units $u_1,\ldots,u_n$ (with $n\in\mathbb{N}$) in a lane and that unit $u_i$ will force $R_i$ buffering units in the same lane to perform undesired removals. Then, the number $R_{n+1}$ for a new entering unit $u_{n+1}$ is determined by comparing the priorities $p(u_i)$ of the units $u_i$. The priority essentially reflects the urgency of a unit to be transported to its destination (see Sect.~\ref{sec:over:ini}). With increasing priority, the necessity of transport to the next node becomes larger.

The number $R=R_{n+1}$ represents the removal index of the lane and is calculated by means of the following iterative formula:

\begin{subequations} \label{eq:removal}
	\noindent
	For the first unit $u_1$ we have
	\begin{equation} 
		R_1 = 0.
	\end{equation}
	If $p(u_i)<p(u_{n+1})$ for all $i=1,\ldots,n$, then  
	\begin{equation} 
		R_{n+1} = n.
	\end{equation}
	If there exists an index $k\in\{1,\ldots,n\}$ with 
	\[
		p(u_k)\ge p(u_{n+1}) \text{ and } p(u_i)<p(u_{n+1}) \text{ for all } i=k+1,\ldots,n,
	\]	
	then we set
	\begin{equation} 
		R_{n+1} = R_k+n-k.
	\end{equation}
\end{subequations}

If a unit that will potentially enter a lane $l$ has the smallest priority of the units in $l$ and all units in $l$ are descending sorted by their priority, no relocation cycles will be necessary, and the removal index $R(l)$ for $l$ is zero. However, the more the actual sequence of the units’ priorities in a lane deviates from the scheduled sequence the units should arrive at their destination, the larger the number of undesired removals. Then the removal index increases. If the unit $u_{n+1}$ that will potentially enter the lane $l$ has the highest priority, then all $n$ currently buffered units in the lane must exit, i.e. the removal index has its maximum value ($R(l)=n$).

Hindrances are not necessarily present in a highly occupied lane, if the newly entered unit and the units buffered in the lane have the same destination without any overlap in the expected production times according to the schedule. If the actual entrance and scheduled exit sequence differ from each other, there will be a direct correlation between the occupancy of the lane and the removal index. The higher the occupancy of the lane, the more relocation cycles are expected.

According to our approach, each unit tries to find a path as hindrance-free as possible. The possible hindrances within a lane $l$ are described by the so-called \textit{hindrance coefficient} $r(l)$, which is calculated from the removal index $R(l)$ according formula (\ref{eq:removal}) and the buffer level of $l$. The hindrance coefficients allow the unit to choose a favorable path to its destination. For this, it is necessary to facilitate an indirect communication between the units. Social insects establish this by the pheromone field \cite{Dorigo2004,Dorigo2000,Theraulaz1999a}. In our model, the hindrance coefficient plays a similar role. It is determined by the occupancy and the removal index of the lane, which influences path finding. In this way, foresighted action of the units is possible.


\subsection{Path Finding with Cycling Capability and Automatic Determination of the Hindrance-Minimal Buffer}
\label{sec:over:search}

Our path finding procedure is based on the A*-algorithm, which extends Dijkstra's algorithm by a destination-oriented heuristic (informed search strategy \cite[p.~94]{Russell2002}). The goal of the path finding is a hindrance-minimal path from the current position of the unit to the destination. Each search must fulfill the temporal restriction given by the scheduled arrival at the workstation.

The best path of a unit is found by the simultaneous consideration of a time and a hindrance component, which defines the temporal urgency to reach its goal on time. In this way, the \textit{hindrance-minimal buffer area} is automatically determined. 

All elements of the transport and buffer system predict their cycle times by means of the exponential smoothing (see Sect.~\ref{sec:over:ct}). The path finding procedure then determines the transport time for a path to the destination based on the cycle times of its nodes and edges.

In the original A*-algorithm the selection of the shortest path takes place on the basis of weighted edges. Since the estimation of possible hindrances is an integral part of our path finding procedure, the evaluation must be extended from a time-based to a more general assessment of utility: As the optimal buffer, if a unit is expected to arrive on time, is the hindrance-minimal buffer, the evaluation must also consider a hindrance component $b_\text{hin}$ besides the time component $b_\text{time}$ (see Fig.~\ref{fig:grob}). Weighting the component $b_\text{hin}$ with a parameter $\beta_\text{path}\ge0$, the assessment of a node $n$ is based on the function
\begin{equation} \label{eq:bewert}
	b_\text{time}(n)+\beta_\text{path}\cdot b_\text{hin}(n) \, .
\end{equation}	

Obviously, the orientation at arrival times and the effort to avoid hindrances can contradict each other. Therefore, a balance between both goals must be found. In principle, an urgent unit needs a fast path and will give little consideration to obstructions of other units. However, if a unit has sufficient time to reach its destination, the path finding selects a path minimizing hindrances of units with higher priority. The relative strength of time orientation and hindrance avoidance decides, whether the units show cooperative or egoistic behavior. 

Our path finding with \textit{cycling capability} permits also paths with cycles. Therefore, a unit can potentially enter the track of a transfer car or a buffer area another time on its path to the destination. This is particularly meaningful, if hindrances are considered in addition to the cycle time in the assessment of alternative paths. 

The cycling capability allows for a deviation from the fastest path, if a hindrance-minimal node can be reached. Sometimes, however, paths with cycles are even time-optimal, if they bypass existing hindrances efficiently.


\subsection{Movement Initiative}
\label{sec:over:ini}

\begin{figure}[htb]
	\centering
		\includegraphics[width=0.95\textwidth]{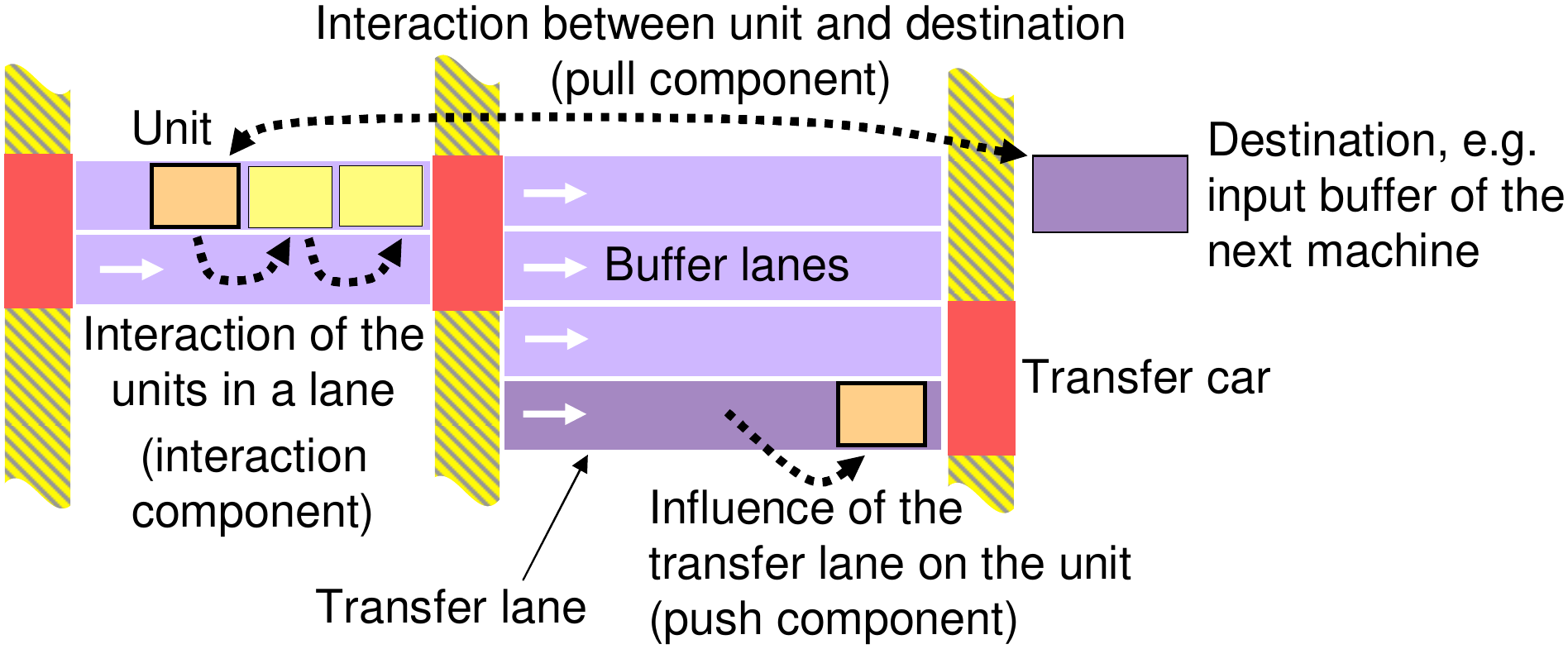}
	\caption{Illustration of the movement initiative and of the interactions, considering decisions at the exit.}
	\label{fig:ini}
\end{figure}

After completed path finding, the ``movement initiative'' decides about the transport to the next buffer area or buffering in the current lane (see Fig.~\ref{fig:ini}). For this, the priority $p(u)$ of the unit $u$ is determined (see Sect.~\ref{sec:over:hind}), considering the following evaluation components: 
\begin{enumerate}
	\item \textit{Pull component} $p_\text{pull}(u)$: 
		\begin{itemize}
			\item The attraction of the destination in order to be on time, considering the predicted cycle time to the unit's destination determined by the path finding procedure (see Sect.~\ref{sec:over:search}) and the scheduled arrival time (i.e. the temporal urgency as a function of transport time and arrival time), 
			\item the attraction of the hindrance-minimal buffer area on the path to the destination.
		\end{itemize}
	\item \textit{Push component} $p_\text{push}(u)$: The repulsive force, if the unit is on a transfer lane that is exclusively intended for transport. 
	\item \textit{Interaction component} $p_\text{inter}(u)$: The interaction with the other units in the lane, taking into account the removal priorities. 
\end{enumerate}

\noindent 
If a unit decides to exit a lane, but is blocked by other units, it informs these about its priority by means of the interaction component. Due to the interaction between the units, the blocking units will react, and exit the lane for the purpose of relocation cycles and a hindrance-free exit of the higher prioritized unit in the lane. 

Let us assume $n$ units $u_1,\ldots,u_n$ in the lane $l$ with the unit $u_1$ at the exit of $l$. The exit of unit $u_1$ will be decided depending on the priority $p(u_1)$ and its components $p_\text{pull}(u_1)$, $p_\text{push}(u_1)$ and $p_\text{inter}(u_1)$. 

The priority $p_\text{lane}$ is a transferred priority. If there is a unit $u$ in another lane that would like to relocate to lane $l$, then $p_\text{lane}=p(u)>0$; otherwise $p_\text{lane}=0$. The interaction component $p_\text{inter}(u_1)$ is calculated by means of
\[
	p_\text{inter}(u_1) = 
		\begin{cases}
			p_\text{lane} & \text{if }n=1,\\ 
			\max\bigl\{p_\text{lane},\, p_\text{pull}(u_2),\, \ldots,\, p_\text{pull}(u_n)\bigr\} & \text{if }n\ge2,
		\end{cases} 
\]
and represents the priorities of the blocked units. The priority $p(u_1)$ is determined by its components according to
\[
	p(u_1) = \max\bigl\{p_\text{pull}(u_1) + p_\text{push}(u_1),\, p_\text{inter}(u_1)\bigr\}.
\]
The push and pull components are summarized, as they do not express the hindrance of other units. Since the interaction component $p_\text{inter}(u_1)$ represents the priorities of the blocked units, it independently influences the determination of the priority $p(u_1)$.

Depending on the priority and its components in comparison with a given \textit{decision threshold} $D\ge0$, the unit $u_1$ will exit, if one of the following conditions is fulfilled:

\begin{enumerate}
	\item Unit $u_1$ decides itself for a removal if \\[1ex]
	$p(u_1)\ge D$ \;\textbf{and}\; $u_1$ can enter a subsequent lane.
	\item A removal is enforced by the handling transfer lane or by obstructed units in the same lane if
	\begin{itemize}
		\item[a)] \; $p_\text{push}(u_1)\ge D$ \quad \textbf{or}
		\item[b)]	\; $p_\text{inter}(u_1)>p_\text{pull}(u_1)$ \;\textbf{and}\; $p_\text{inter}(u_1)\ge D$.
	\end{itemize}
\end{enumerate}

As soon as a unit that has requested removals is not hindered anymore, it will call the transfer car in order to exit lane $l$ and the best subsequent lane to enter the next buffer area will be selected. If non of the conditions is fulfilled, the unit $u_1$ (and all blocked units) will wait. 

With increasing value of $D$, the unit decides to exit later, i.e. $D$ represents the reactivity of the unit to external events\footnote{More details about this decision-making process can be found in Appendix C.3.5 of Ref. \cite[p.~188]{Seidel2007}.}.


\subsection{Hindrance-Avoiding Selection of the Next Lane} 
\label{sec:over:select}

If the movement initiative triggers the exit of a unit, a selection of the most suitable lane in the next buffer area is needed. Of course, it would be favourable, if the units were buffered in the same sequence in which they are supposed to exit the lane according to the (optimized) production schedule. Thus, the best lane is the one whose sequence of units differs as little as possible from the scheduled order of units to exit. To occupy the transfer car as little as possible, a unit exits its lane only after the following lane has been chosen. This guarantees that there will be empty capacity for the unit in the selected lane when it actually reaches this lane. Altogether, the selection process has to estimate the suitability of the lanes \textit{and} to examine the availability of sufficient buffer capacity.

\begin{figure}[bth]
	\centering
	\includegraphics[width=1.0\textwidth]{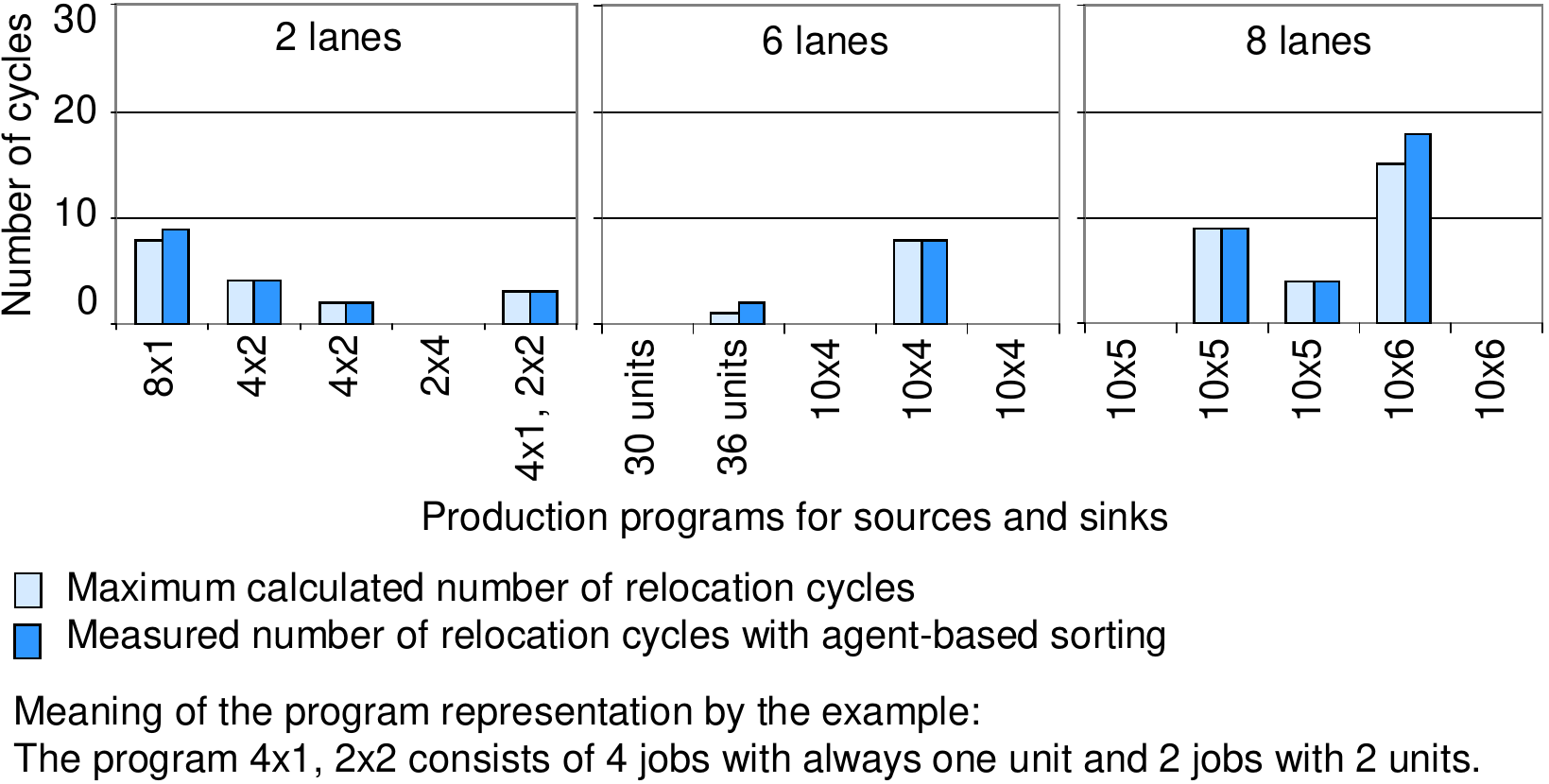}
	\caption{Comparison of the estimated number of necessary relocation cycles and the actually measured number generated by our agent-based sorting.}
	\label{fig:zyklen}
\end{figure}

The selection process is made via \textit{agent-based sorting}, i.e. each unit acts as an autonomous agent and selects the lane independently. There is no central decision maker, who performs the sorting. From the point of view of an entering unit, the sequence of the units in the lane is compared with the scheduled order of exits (see Fig.~\ref{fig:zyklen}). Note, however, that a binary interchange of units is not possible, in contrast to most conventional sorting procedures.

\noindent
The selection process consists of the following steps:
\begin{enumerate}
	\item If the buffer area consists of only a single lane, then this lane is selected.
	\item If there is more than one lane in the area and if there exists a lane with a unit of the same job waiting at the last position of that lane for the same workstation, then decide for this lane.
	\item Otherwise, the lane selection involves the following components:
		\begin{itemize}
			\item \textit{Hindrance component $b_\text{hin}$}: For all lanes, the suitability of entering is evaluated. This considers all hindrances, which the unit has to expect \textit{and} which it may cause. The smaller the evaluation $b_\text{hin}(l)$ of the lane $l$, the greater the suitability of that lane. The definition of the function $b_\text{hin}$ determines the quality of the agent-based sorting. A feasible function $b_\text{hin}$ is described in Ref. \cite[p.~191]{Seidel2007}.
			\item \textit{Resource component $b_\text{res}$}: Since the lanes can have different widths and the units can have different space requirements, the lanes are evaluated with respect to the utilization of the provided buffer capacity. A possible evaluation function is $b_\text{res}(l)=w(l)/w_\text{unit}$ with the width $w(l)$ of the lane $l$ and the width $w_\text{unit}$ of the deciding unit.
		\end{itemize}
		Finally, $b(l)=b_\text{hin}(l)+b_\text{res}(l)$ is calculated for all lanes $l$ of the buffer area.
		A lane $l_0$ is selected when it fulfills $b(l_0)=\min\limits_{l:\,w(l)\ge w_\text{unit}} b(l)$.		
\end{enumerate}


\section{Mathematical Abstraction of Interdependencies in the Transport and Buffer System}
\label{sec:steps}

\begin{figure}[htb]
	\centering
	\includegraphics[width=1.0\textwidth]{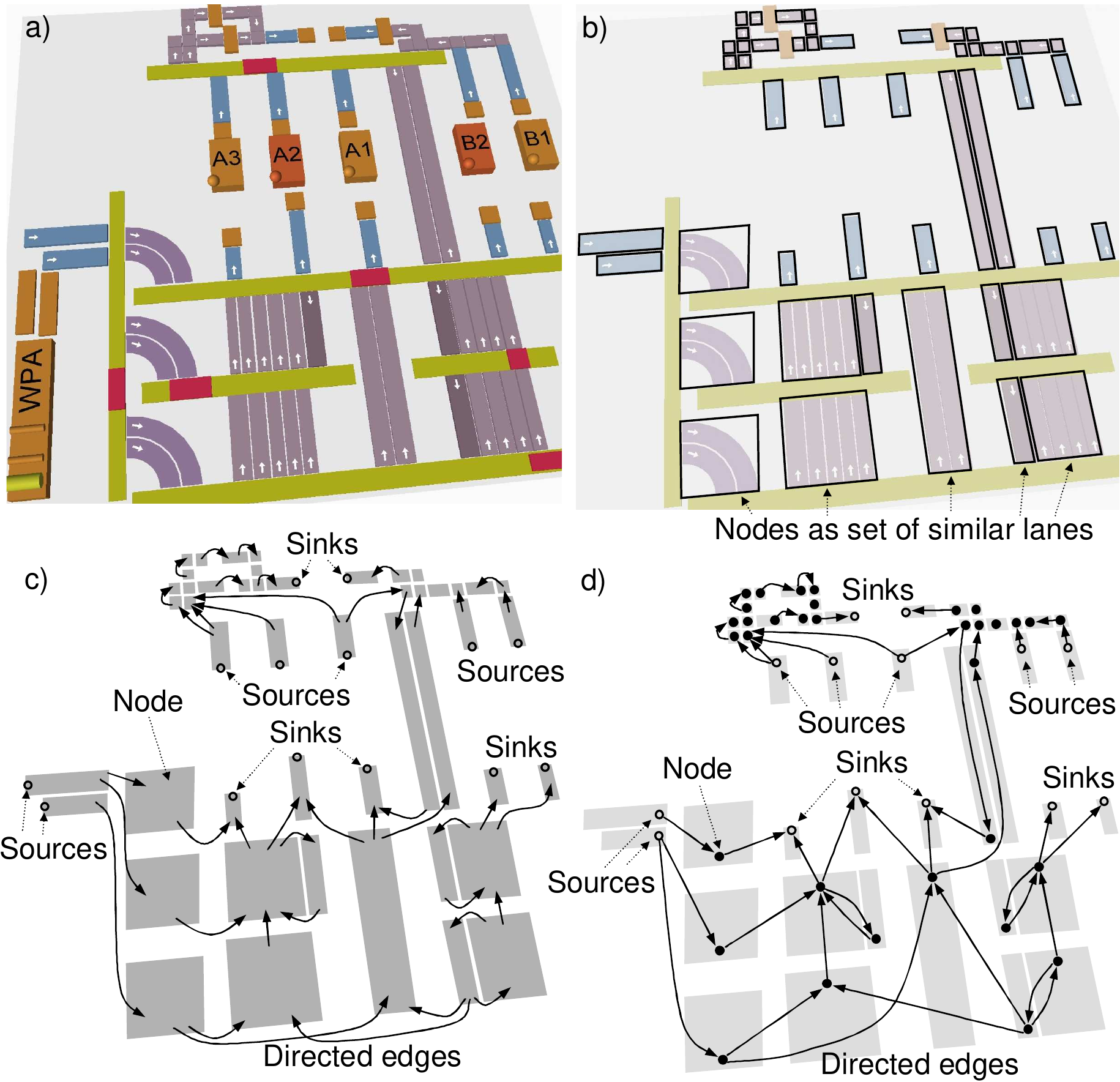}
	\caption{Different steps in the abstraction of a factory layout. a) Example of a 
plant layout.  b) Logical representation of the buffer system as a set of nodes of a mathematical graph. c) Representation of the transfer cars and the dispatch machines as directed edges of the graph (only a subset of edges is represented). d) Resulting mathematical graph representing the factory layout.}
	\label{fig:netzwerk}
\end{figure}

In our agent-based approach, the layout of a plant of, for example, a packaging manufacturer, is represented by a mathematical graph G with nodes and directed edges \cite{Ahuja1993, Godsil2001}. Figure~\ref{fig:netzwerk} shows the subsequent steps in the abstraction of a factory layout. 

\begin{figure}[htb]
	\centering
		\includegraphics[width=1.0\textwidth]{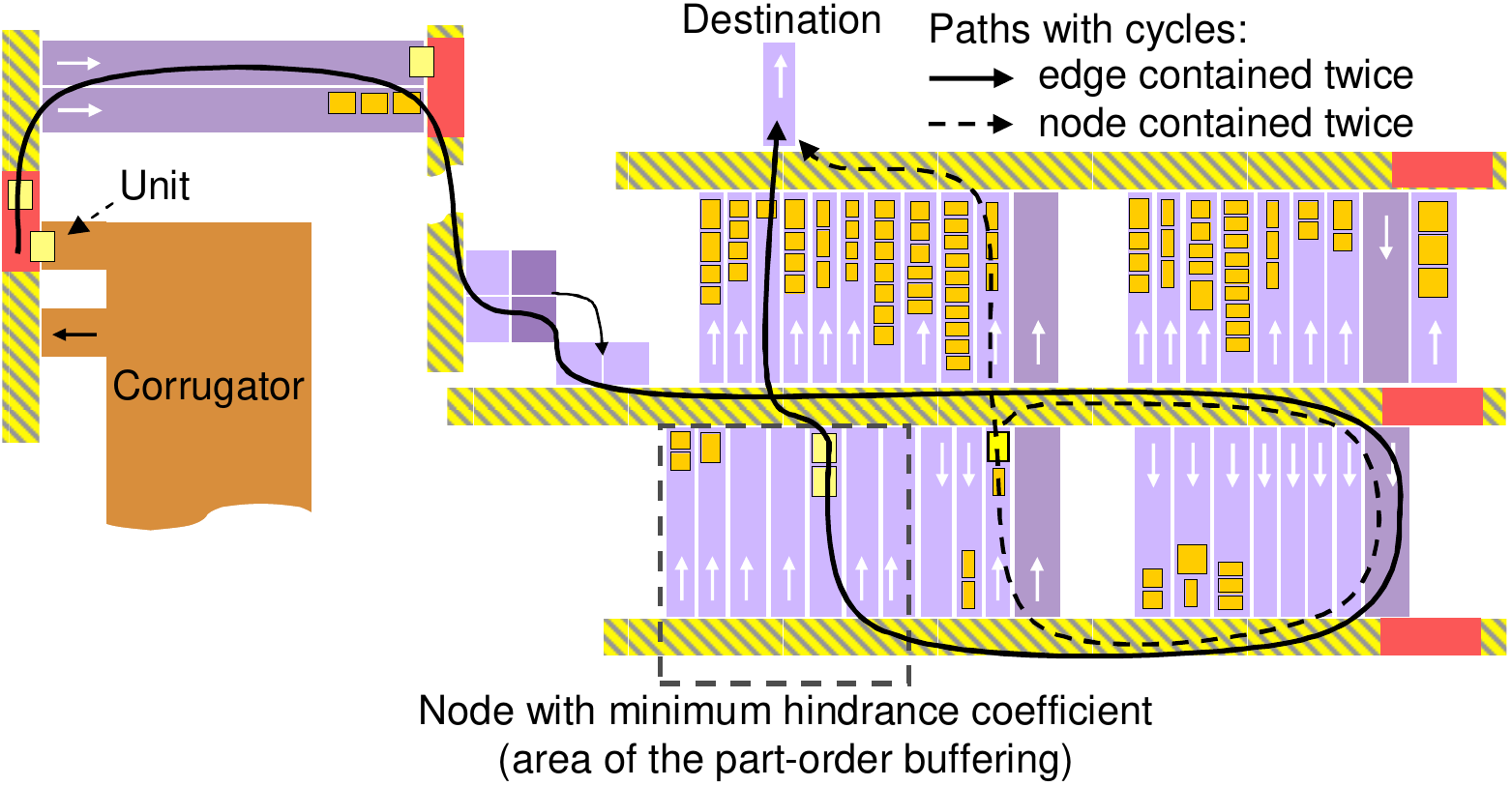}
	\caption{Illustration of two paths containing the same edge twice (solid line) or the same node twice (dashed line). Cycles are generated by selecting a hindrance-minimal buffer area (i.e. the node with smallest hindrance coefficient).}
	\label{fig:bsp2}
\end{figure}

Note that a  unit can use \textit{any} lane of a buffer area on its path (e.g. the area of the part-order buffering in Fig.~\ref{fig:bsp2}): When searching for a path to the destination, the identification of an available link is more important than the determination of the respective lane.
Therefore, homogeneous, parallel lanes are combined into sets of lanes, which form the nodes of the graph $G$ (see Fig.~\ref{fig:netzwerk}b). Since turntables do not have a given direction, each of them is an individual node in the graph $G$.

The track of a transfer car connects different lanes. Similarly, the dispatch machines (for example pallet inserters) connect several lanes, since they have an input and an output buffer, and the number of units is not changed while being processed, i.e. the machines transport an individual unit from the input buffer to the output buffer. Both, a dispatch machine and a transfer car track can be regarded as linkage of two nodes, i.e. they form the edges in the graph. Since only directed edges are modeled, a bidirectional transfer (e.g. between two neighboring turntables) is represented by two oppositely directed edges (see Fig.~\ref{fig:netzwerk}c).

The remaining workstations (e.g. the corrugator and the converting machines) form new units, so that the number of the ingoing and outgoing units can be different. Therefore, these machines are modeled as sources and sinks respectively, which are typically interconnected in a certain way. The stacker of the corrugator forms stacks of raw boards for the output buffer and can be regarded as source. The prefeeder of a converting machine takes the units from the input buffer and, therefore, is a sink. The load former creates stacks with boards and can be interpreted as source (like the stacker of the corrugator). Finally, we have the exits of the plant, which are sinks. The corresponding mathematical graph $G$ represents the interaction of the different material handling elements of a plant in our agent-based simulation (see Fig.~\ref{fig:netzwerk}d).


\section{Description of the Movement of Units in the Modeled System}
\label{sec:desc}

\begin{figure}[htb]
	\centering
		\includegraphics[width=1.0\textwidth]{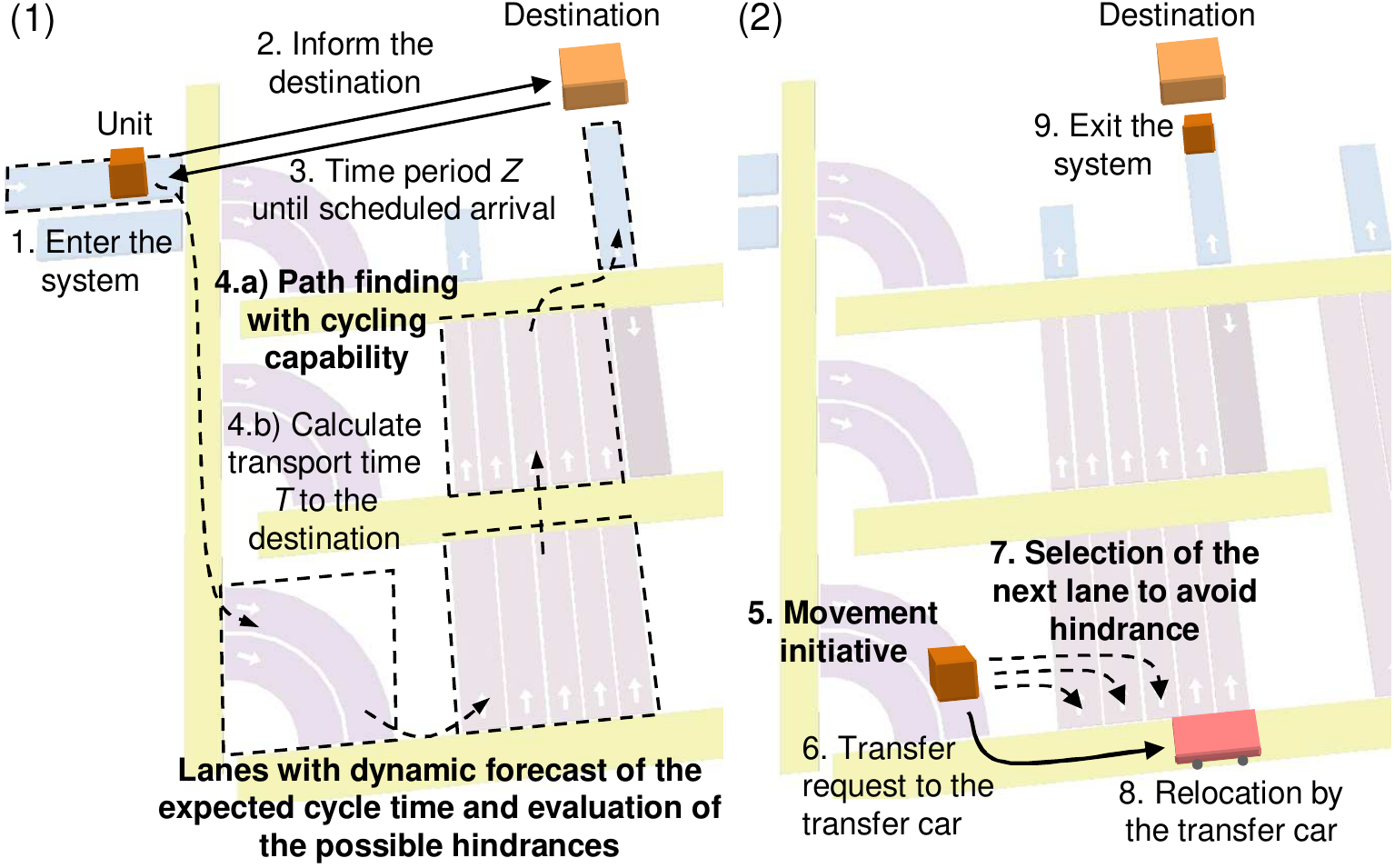}
	\caption{Causal operational sequence of the movement of units in the modeled transport and buffer system.}
	\label{fig:fein1}
\end{figure}

Figure~\ref{fig:fein1} summarizes the different procedures contributing to the definition of the movement of a unit from its source to its sink (destination). 

\begin{enumerate}
	\item The movement procedure starts with the leaving of a workstation and with entering its output buffer. Consequently, the unit enters the transport and buffer system at this time. 
	\item The unit transfers the information that is has entered the system to the destination. 
	\item The destination schedules the sequence of the units according to the production program, in which they can enter the input buffer of the machine. From this, the destination derives their expected order of arrival, using the expected transport time $T$ from their current position to the destination. 
The scheduled arrival time $Z$ is then transferred back to the units. If $T\le Z$, the unit has enough time to arrive at the destination on time and will buffer at a node for a time period $Z-T$. If, however $T<Z$, the unit will probably not arrive on time and will get high priority.
	\item Considering time period $Z$ until the scheduled arrival at its destination, the unit determines the best path from the current position to the input buffer of the machine. Since the nodes of the graph abstracting the factory network may represent \textit{several} homogeneous lanes, the path does not specify the lane at this stage. The new lane is selected, when the unit is transferred to the next node. After the unit has determined its path, it registers itself at the nodes and edges of the path. If the estimated (partial) cycle times at some nodes or edges change, the unit is informed about this. It may then adapt its expected cycle time or determine a new path.
	\item The exit of the unit from the lane is decided according to different criteria 
(see Fig.~\ref{fig:ini}). Blocked units report their priority to the next and eventually to the first unit in the respective lane, so that the blocking units consider exiting. 
	\item If the exit was decided and the unit is not blocked, the transfer car receives a request to relocate the unit to the next node. 
	\item If the next node consists of several lanes, the best lane is selected, considering the lane width and the possible hindrance of units buffered in those lanes. 
	\item As soon as the relocation is completed, the unit enters a lane of the next node, and the path finding procedure starts again.
	\item If the unit arrives at its destination, it leaves the modeled system and finishes its movement procedure.
\end{enumerate}


\section{Implementation of the Model in a Simulation Environment}
\label{sec:impl}

\begin{figure}[htb]
	\centering
	\includegraphics[width=1.0\textwidth]{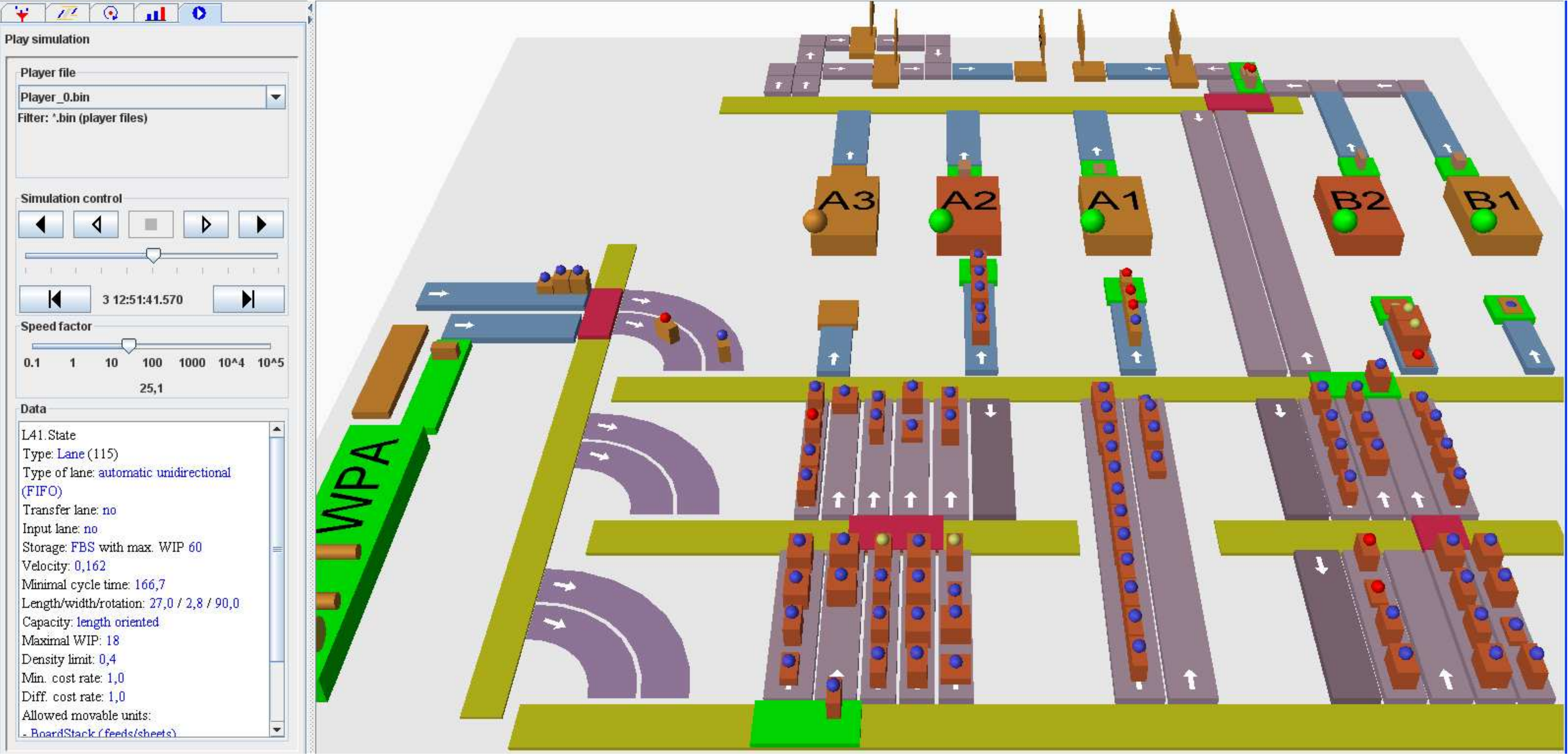}
	\caption{Program interface representing a model of a packaging plant.}
	\label{fig:modeler}
\end{figure}

We have also developed a simulation software for our model of the transport and buffer system.
Our software consists of different modules, which are controlled over a common program interface (see Fig.~\ref{fig:modeler}). The modules are developed as independent software units communicating with each other. 

The simulator contains a library for the simulation of discrete events. During the simulation, the behavior of the objects is recorded and passed on to the statistics module, which automatically generates a HTML page with the simulation results. The units and their spatiotemporal dynamics are visualized with the help of a simulation player. Additionally, variables characterizing the units and the production machines can be displayed in separate windows.

\section{Path Finding as Basis for the Interaction of Units}
\label{sec:res}

\begin{figure}[htb]
	\centering
	\includegraphics[width=0.8\textwidth]{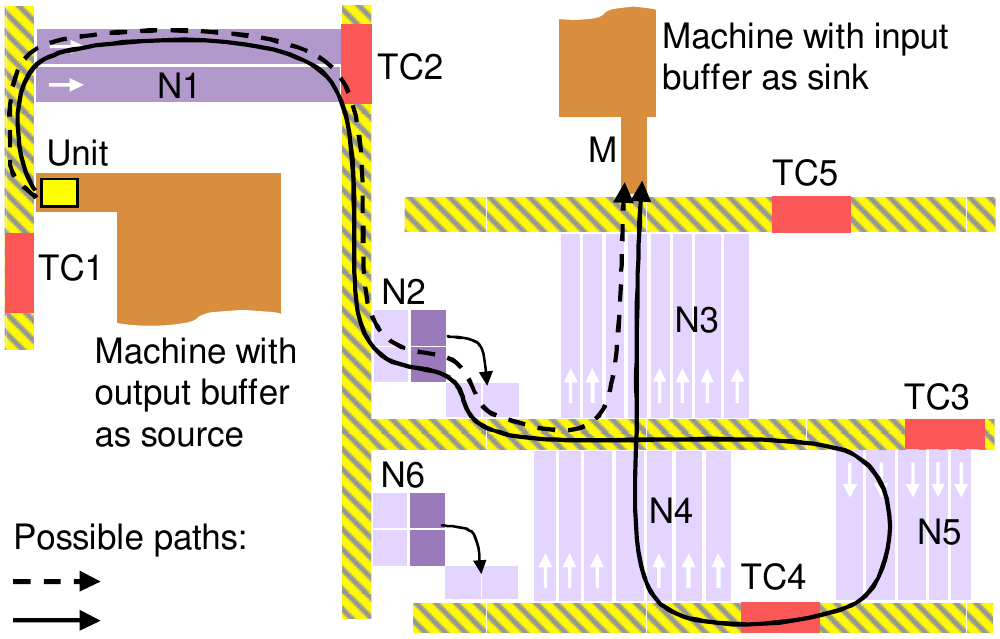}
	\caption{Plant layout as basis for the choice of alternative paths.}
	\label{fig:bsp3}
\end{figure}

For the following analysis, we will assume the plant layout shown in Fig.~\ref{fig:bsp3}. Our goal is to study, whether our path finding algorithm produces reasonable results. We will start investigating the influence of certain parameters on a \textit{single} unit. Afterwards, we will continue with an analysis of the \textit{interaction} of units belonging to \textit{different} production jobs.

\subsection{Deviation from the Fastest Path by Variation of the Weight Parameter $\beta_\text{path}$}
\label{sec:res:path}

According to formula (\ref{eq:bewert}), the evaluation function of a node in the path finding algorithm consists of a time component and a hindrance component, which is weighted by the parameter $\beta_\text{path}$. With $\beta_\text{path}=0$, only time-oriented goals are considered during the evaluation. The possibility of the deviation from the fastest path is reached by a parameter $\beta_\text{path}> 0$. Let us assume that the time until the scheduled arrival is $Z=1400$~s and that the model parameters are specified 
as listed in the following table\footnote{The hindrance $c(n)$ of the elements $n$ listed in the table is calculated according to $c(n)=t(n)\cdot r(n)$, apart from the input buffer M, for which the hindrance is set to 0, as the units enter the input buffer already in a sorted way.}:

\begin{center}
\begin{tabular}{|@{\hspace{2mm}}l@{\hspace{2mm}}||@{\hspace{2mm}}c@{\hspace{2mm}}|@{\hspace{2mm}}c@{\hspace{2mm}}|@{\hspace{2mm}}c@{\hspace{2mm}}|}
\hline
   Element $n$ & Cycle time $t(n)$ & Hindrance          & Hindrance $c(n)$ \\
               &                   & coefficient $r(n)$ & \\
\hline
Nodes N1, N2,  &      &              &     \\
N3, N6         & 70~s &          1.5 & 105 \\
Node N4        & 70~s & \textbf{1.2} &  84 \\
Node N5        & 70~s & \textbf{1.0} &  70 \\
Input buffer M & 70~s &          2.0 &   0 \\
Edges          & 30~s &          2.5 &  75 \\
\hline
\end{tabular}  
\end{center}

\begin{figure}[htb]
	\centering
	\includegraphics[width=1.0\textwidth]{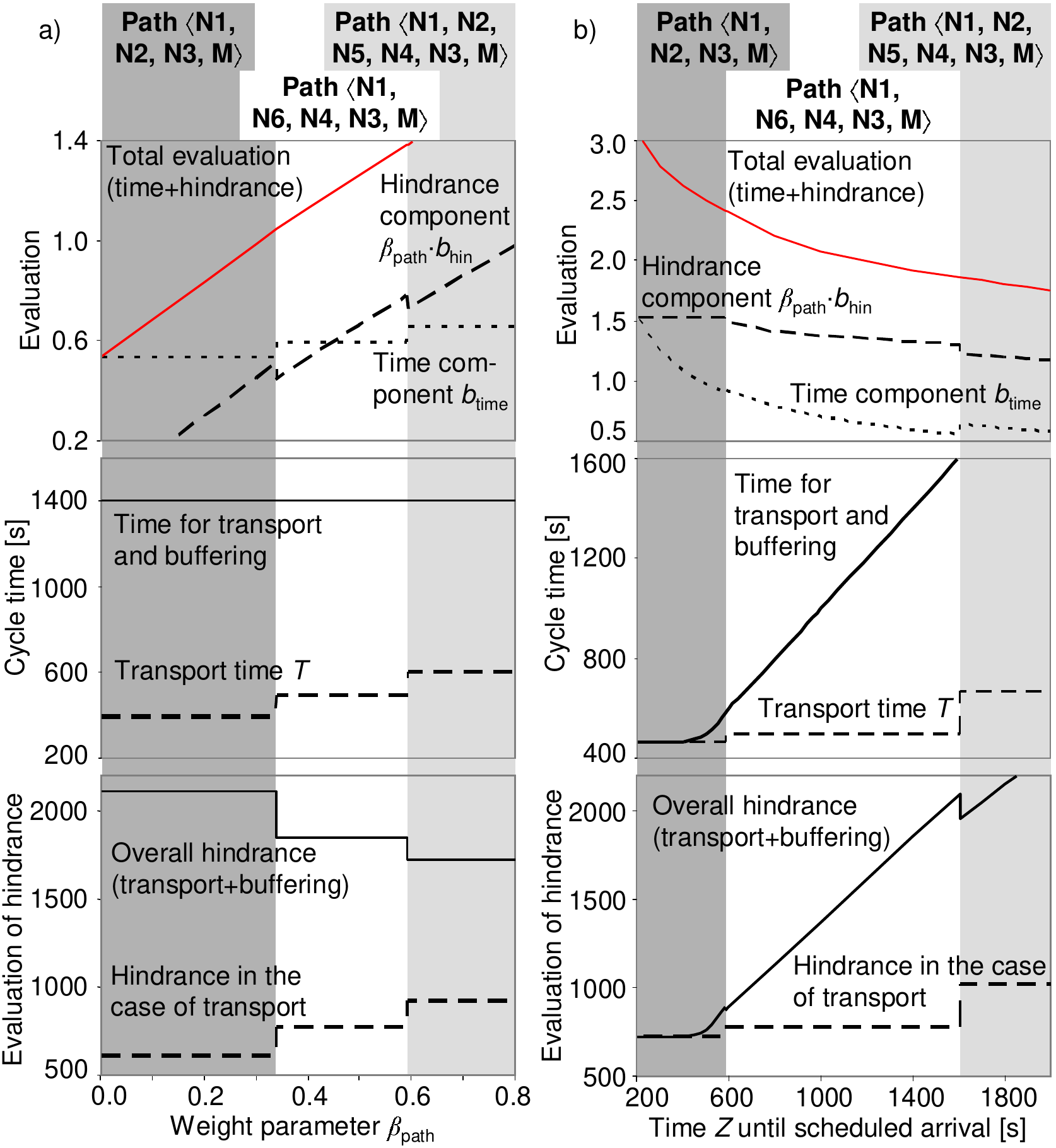}
	\caption{Influence on the route choice of a) the weight parameter $\beta_\text{path}$ and b) the time $Z$ until the scheduled arrival of the unit at the input buffer M of the destination.}
	\label{fig:bsp3a}
\end{figure}

Figure~\ref{fig:bsp3a}a shows the simulation results as a function of $\beta_\text{path}\ge0$. The upper diagram shows the evaluation function with its time and hindrance components. The values of these components are only determined by the respective path. They do not vary with the weight $\beta_\text{path}$, while the weighted sum (\ref{eq:bewert}), of course, does. Note, however, that the overall evaluation (\ref{eq:bewert}) is a smooth function, since the transitions from one path to another occurs for a value of $\beta_\text{path}$, at which their overall evaluations cross each other (i.e. where they are identical).

In Fig.~\ref{fig:bsp3a}, the transport time $T$ and the cycle time for transport and buffering are represented in the diagram in the middle. The cycle time contains a buffering time, if the unit arrives at the destination too early. Therefore, the cycle time is constant and equals $Z=1400$~s as long as all three paths require lower transport times than given by this value. The transport time $T$ changes only, when transitions to another path occur.

The lower diagrams in Fig.~\ref{fig:bsp3a} show the evaluation of the hindrance expected during the transport. Note that, although the transport time $T$ becomes larger, the overall hindrance reflecting transport \textit{and} buffering decreases at the transition points. 

The simulation results presented in Fig.~\ref{fig:bsp3a} illustrate that the path with the shortest transport time is chosen for small values of $\beta_\text{path}$. Note, however, that
nodes N4 and N5 have lower hindrance coefficients than the other ones. Consequently, these nodes become more attractive for larger weights $\beta_\text{path}$, and the hindrance component becomes more influential on the path finding procedure as $\beta_\text{path}$ increases. Therefore, 
nodes N4 and N5 are integrated into the path for sufficiently large values of $\beta_\text{path}$, 
although this leads to increased transport times (see the white and light grey areas).


\subsection{Deviation from the Fastest Path by Variation of the Time Period $Z$ Until the Scheduled Arrival at the Destination}
\label{sec:res:Z}

Another important factor of path choice is the time period $Z$ until the scheduled arrival at the destination. For small $Z$, the destination can possibly not be reached on time, so that a fast path is selected. The larger $Z$, the more likely a hindrance-avoiding path is chosen\footnote{That applies only to the assumption of $\beta_\text{path}>0$, so that the hindrance component is actually considered.}. 

For our analysis, we have assumed the weight $\beta_\text{path}=1$ and model parameters according to the following table:

\begin{center}
\begin{tabular}{|@{\hspace{2mm}}l@{\hspace{2mm}}||@{\hspace{2mm}}c@{\hspace{2mm}}|@{\hspace{2mm}}c@{\hspace{2mm}}|}
\hline
   Element $n$ & Cycle time $t(n)$ & Hindrance coefficient $r(n)$ \\
\hline
Nodes N1, N3, N6 &           70~s & 1.5          \\
Node N2          & \textbf{140~s} & 1.5          \\
Node N4          &           70~s & \textbf{1.2} \\
Node N5          &           70~s & \textbf{1.0} \\
Input buffer M   &           70~s & 2.0          \\
Edges            &           30~s & 2.5          \\
\hline
\end{tabular}  
\end{center}

Figure~\ref{fig:bsp3a}b shows the simulation results as a function of the time period $Z$ until the scheduled arrival at the destination. For small values of $Z$, the fastest path is selected with a transport time of $T=470$~s. A transport via the second path $\langle$\,N1, N6, N4, N3, M\,$\rangle$ requires $T=500$~s. The latter path is only selected, when the hindrance during buffering has become significant. First signs of hindrance effects due to buffering can be seen for $Z > 470$~s.


\subsection{Blockage}
\label{sec:res:blockade}

The blockage of a node is a further variable influencing path choice. We will show that path finding avoids nodes\footnote{A node can be avoided only, if there are alternative paths.} at which the material flows are in a danger to be blocked.

If the requested removals from a node are not processed, then the cycle time for transport increases even without any additional buffering at the node, just because of a temporary blockage of the units. Therefore, an emerging blockage due to delayed removals of units may be reflected by large transport cycle times.

If the units of a node are in a highly unsorted order, then the blockage can be caused by frequent relocation cycles binding possible removal capacity. Since the possible obstructions are described by the hindrance coefficient, a high value of this coefficient reflects the danger of blockages. 

An \textit{actual} blockage of a node develops either due to delayed removals or due to an increase in the number of hindrances. Therefore, the blockage of a node can be recognized by its large hindrance coefficient and the increasing cycle time for transport (even without additional buffering).

Let us now assume the scheduled arrival time $Z=1050$~s, the weight $\beta_\text{path}=1$ and the parameters listed in the following table:

\begin{center}
\begin{tabular}{|@{\hspace{2mm}}l@{\hspace{2mm}}||@{\hspace{2mm}}c@{\hspace{2mm}}|@{\hspace{2mm}}c@{\hspace{2mm}}|}
\hline
   Element $n$ & Cycle time $t(n)$ & Hindrance coefficient $r(n)$ \\
\hline
Nodes N1, N3   &     70~s & 1.5 \\
Node N2        & variable & variable \\
Node N4        &     70~s & 1.2 \\
Node N5        &     70~s & 1.0 \\
Node N6        &    140~s & 1.8 \\
Input buffer M &     70~s & 2.0 \\
Edges          &     30~s & 2.5 \\
\hline
\end{tabular}  
\end{center}

\begin{figure}[htb]
	\centering
	\includegraphics[width=1.0\textwidth]{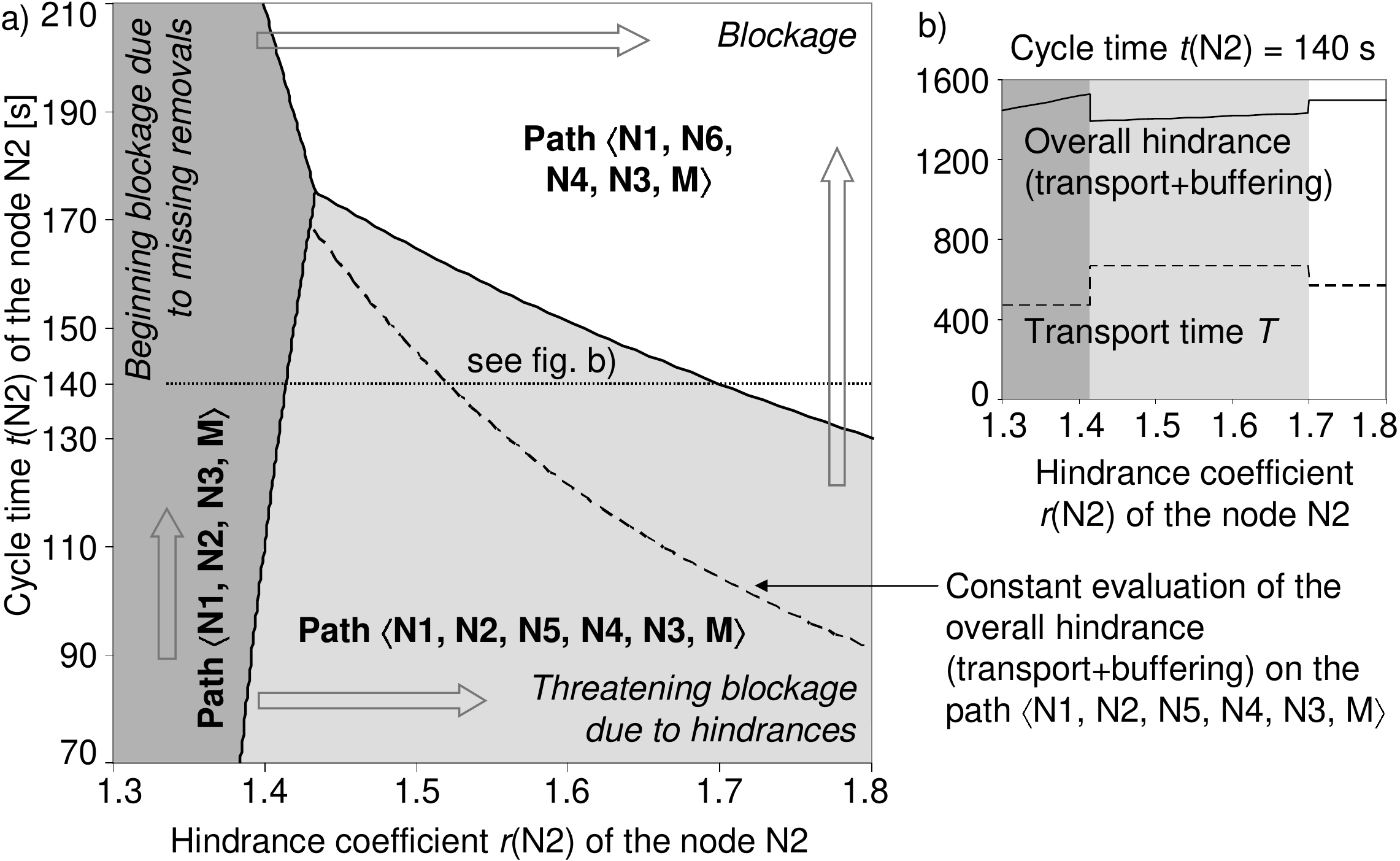}
	\caption{a) Path finding as a function of cycle time $t$(N2) and hindrance coefficient $r$(N2) of node N2. The dashed line corresponds to the isocline of constant hindrance. b) Overall hindrance and transport time $T$ along the isocline $t({\rm N2})=140$~s as a function of the hindrance coefficient $r$(N2). One can see that the overall hindrance may drop, while the transport time $T$ increases.}
\label{fig:bsp4a}
\end{figure}

Figure~\ref{fig:bsp4a}a shows the route choice as a function of the cycle time $t$(N2) and hindrance coefficient $r$(N2) of node N2. For small hindrance coefficients, the path $\langle$\,N1, N2, N3, M\,$\rangle$ is always selected, since it is fastest. 

However, for cycle times $t$(N2) smaller than 130~s, a transition to the path $\langle$\,N1, N2, N5, N4, N3, M\,$\rangle$ containing the hindrance-minimal node N5 takes place, when the coefficient $r$(N2) grows.

If the regime of blockage of node N2 is finally reached, there is always a transition to the path $\langle$\,N1, N6, N4, N3, M\,$\rangle$, which does not contain the node N2. Thus, the blockage of N2 is recognized and N2 is avoided. Whether a node is recognized as blocked, depends on the following: 

\begin{itemize}
	\item the intensity of the blockage, which is determined via the cycle time (without buffering) and the hindrance coefficient of the node, 
	\item parameters such as cycle times or hindrance coefficients characterizing the efficiency of the infrastructure of the plant, 
	\item the time period $Z$ to the scheduled arrival at the destination, and 
	\item the weight parameter $\beta_\text{path}$.
\end{itemize}


\subsection{Cooperative Versus Egoistic Behavior}
\label{sec:res:coop}

\begin{figure}[htb]
	\centering
	\includegraphics[width=0.75\textwidth]{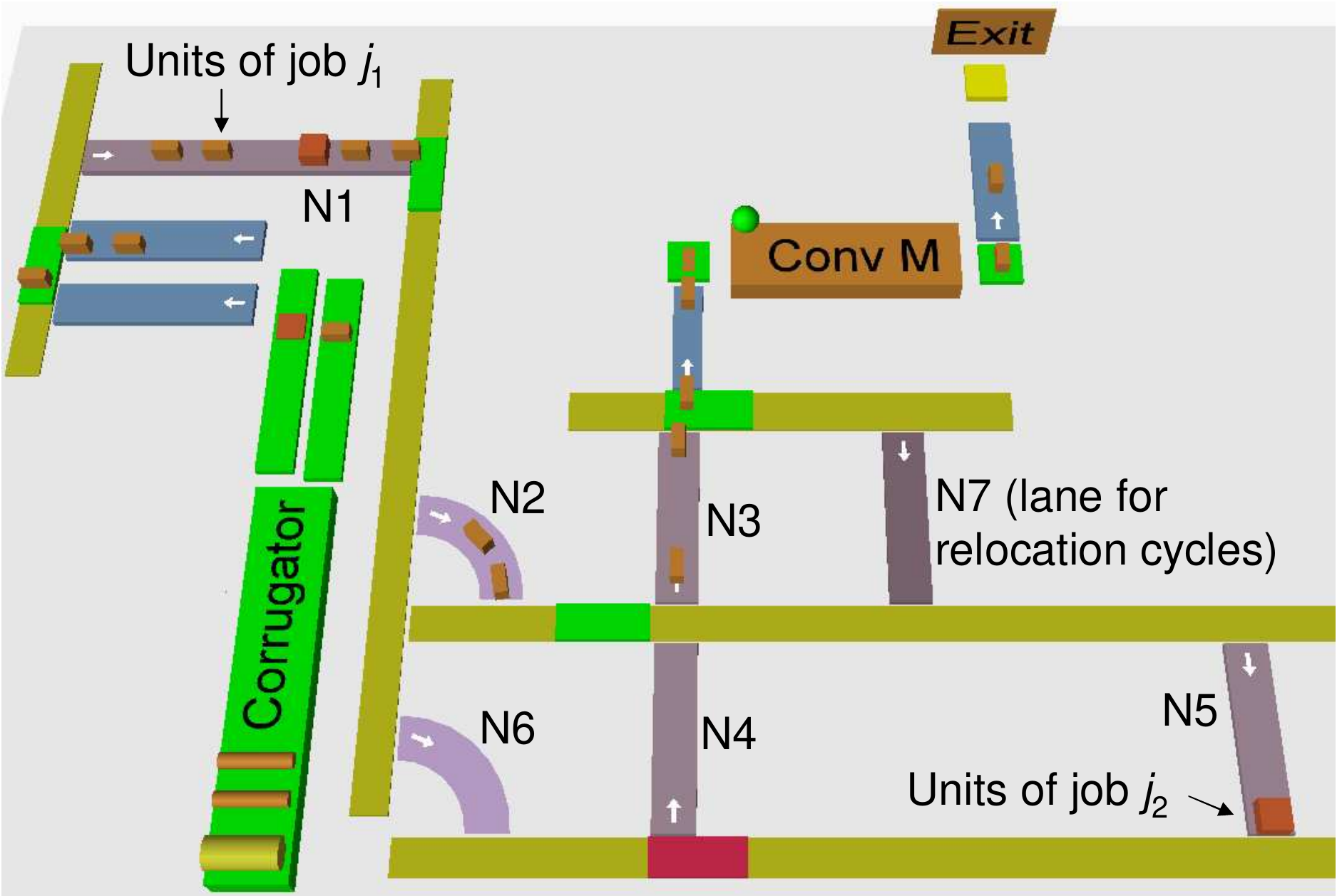}
	\caption{Plant layout for the simulation of ``social'' (cooperative) behavior.}
	\label{fig:bsp5}
\end{figure}

Let us now simplify the plant layout of Fig.~\ref{fig:bsp3} as depicted in Fig.~\ref{fig:bsp5}. The machine ``Corrugator'' produces intermediate units (boards) of two jobs. Job $j_1$ has 100 units with a processing time of 1 min for one unit. The second job $j_2$ contains 4 units with a processing time of 25 min per unit. The machine ``Conv M'' converts\footnote{Typical converting processes are cutting, printing, creasing and gluing \cite{SCAAylesford}.} the boards to finished packages that leave the plant at the station ``Exit''. The converting machine is supposed to complete the units belonging to jobs $j_1$ first, and afterwards the units of job $j_2$. 

\begin{figure}[htb]
	\centering
	\includegraphics[width=1.0\textwidth]{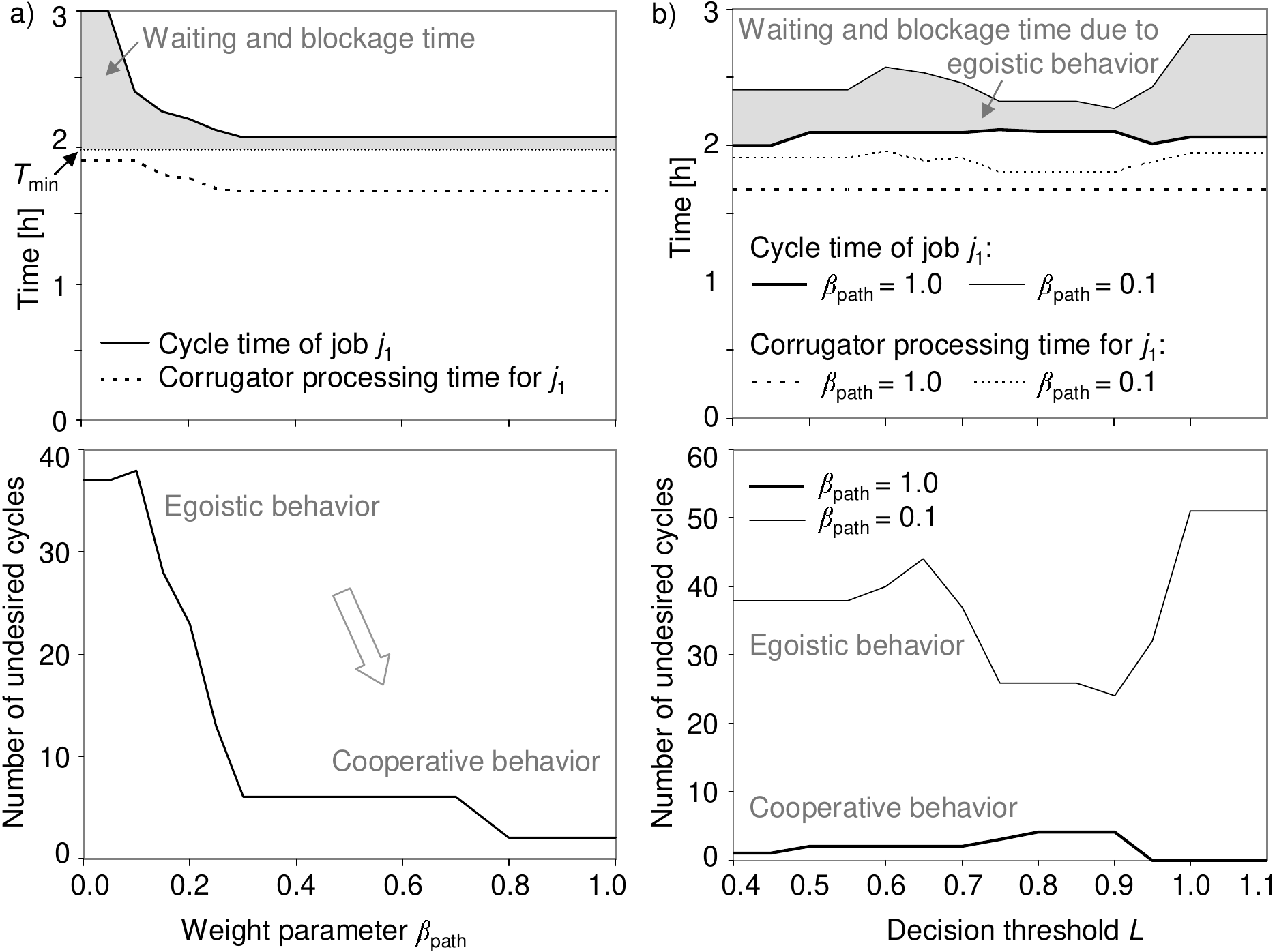}
	\caption{a) The parameter $\beta_\text{path}$ is used to induce cooperative behavior. b) Influence of the decision threshold $L$ (see Sect.~\ref{sec:over:ini}).}
	\label{fig:bsp5a}
\end{figure}

If the corrugator produces the jobs $j_1$ and $j_2$ in the same order as they are converted, the minimum cycle time is $T_{\min}=116.25$~min for all units of job $j_1$ from the beginning at the corrugator to the finishing of the last unit at the converting machine\footnote{A time period of about 16~min is needed by the transport of the units between both machines, the waiting of the converting machine, and the setup of the machines.}. The efficiency of the cooperative (``social'') behavior can be quantified by comparing the actual cycle time with the minimum time $T_{\min}$. The weight parameter $\beta_\text{path}$ and the decision threshold $L$ are the factors influencing the (social) behavior of the units\footnote{Actually there is another influencing parameter (see \cite[p.~186]{Seidel2007}), which motivates the unit to move to the hindrance-minimal node of the path. Since only 4 out of 104 units behave egoistically, they are forced to cooperate in the interactions with the others (see Fig.~\ref{fig:ini}).}. 

Now, let the corrugator execute both jobs $j_1$ and $j_2$ concurrently through duplex production (see Ref. \cite{SCA}), i.e. the first unit of job $j_2$ is finished at the same time as the 25th unit of $j_1$. Both jobs need 100 min processing time and will be finished by the corrugator at the same time. The processing order at the converting machine shall remain first $j_1$, then $j_2$. 

As the units of job $j_1$ have to hurry, they have a high priority to get to their destination. Their ``movement initiative'' causes the units of job $j_2$ to consider this. Therefore, the decision threshold
has only an influence on the temporal sequence of the decision-making process. Small variations in time can lead to large variations in the number of cycles. Obviously, this can occur only in the regime of egoistic behavior with many cycles. 

For small values of $\beta_\text{path}$, the units of job $j_2$ select the fastest path to the destination without consideration of the possible obstruction of units belonging to job $j_1$. This causes removals\footnote{The removals lead to relocation cycles via N7.} of units belonging to job $j_2$ and leads to many relocation cycles (see Fig.~\ref{fig:bsp5a}a). As the units of $j_2$ hinder the units of $j_1$, they show egoistic behavior. However, with increasing $\beta_\text{path}$, the units of $j_2$ consider hindrances and decide for a buffering at node N5. By this cooperative behavior of the units of job $j_2$, the hindrance of the units of the more urgent job $j_1$ is avoided.


\subsection{General Characteristics}

Although our results were obtained for special plant layouts of a packaging manufacturer, our findings can be extended to more general settings: Under certain conditions, our model allows the units to diverge from the fastest path to the destination. For example, if a late arrival at the destination is expected, the unit possibly decides for a longer path, if this facilitates buffering in an area with fewer hindrances.

Furthermore, our algorithm concept allows a unit to detect a substantial increase of the expected transport time for the decided path. If the anticipated transport time becomes too high, the unit possibly determines a better path to its destination and decides to bypass a congested buffer area.

In general, the characteristics of our approach reflect coordinated behavior as it shall be found in real plants operated by a central and goal-oriented planner considering reasonable prioritizations. The resulting transport of units ensures the feeding of each workstation with the right product in the right quantity at the right point in time \cite{Gudehus2000, Hopp2000, Nahmias2001}.


\section{Discussion}
\label{sec:disc}

This chapter has described the modeling of transport and buffer systems based on an arbitrary layout and the movement of the units within that system considering the scheduled arrival sequence at the workstations. 

We have abstracted the material handling system as a mathematical graph with nodes and directed edges. Units (representing products or work in process) are treated as agents, which operate on the graph and interact in direct and indirect ways. The goal of their operation is the avoidance of blockages, which is achieved by indirect interactions minimizing the expected hindrances. So, the reduction of hindrances in the system is facilitated by cooperative behavior of the agents.

Our model contains decentralized decisions, which enable a flexible adjustment to the current situation in the plant. In particular, suitable local interactions can avoid mutual hindrances of the units. A combination of local and centralized procedures facilitate the arrival of the units at their destination in the right sequence. On the one hand, the units are arranged in the right order by means of a classical sorting algorithm in accordance with the (optimized) production program. On the other hand, the units are sorted by relocations based on local interactions. 

The high flexibility with respect to the restructuring of the layout or changes in the operation of a production system is a major advantage of the agent-based approach. Not only can new scenarios, such as effects of machine breakdowns or allocations of buffer areas to machines, be easily simulated and quickly evaluated, but also can the effects of newly installed machines or relocated workstations on the operational procedures be efficiently analyzed.
 
The developed simulation software can support planners in plants of packaging and other manufacturers in creating better production programs. Since the effects of the generated programs are simulated in advance, the planner can test, which production programs are expected to cause operational hindrances in the material flows and consequential disturbances in the production, and which ones not.

Note that the decentralized (local) control procedures of our agent-based approach could be also implemented by means of RFID tags attached to the units. Due to its flexibility regarding the layout and operation of production systems, this implementation would be applicable to many different plants. Then, various control strategies could be easily implemented by adjusting a few parameters only, thereby determining different operational programs.

Rather than using RFID tags just to replace classical bar codes, our proposed implementation would enable more flexible, robust, and efficient decentralized control approaches in complex production systems. In our case, the units would search a path through the production plant in an autonomous way, considering the scheduled completion times and hindrances in the system. While performing this task, our agents would use rudimentary intelligence and forecasting capabilities. This would both, generate and use cooperative (``social'') behavior of the individual units.


\subsection*{Acknowledgments}

The authors wish to thank Andrew Riddell, John Williams, and Brian Miller of SCA Packaging Ltd.
for their great support of this study.

%
%
%
\bibliographystyle{abbrv}
\bibliography{lit}
%


\printindex
\end{document}